\documentclass[
  aps,
  pra,
  amsmath,amssymb,amsfonts,
  twocolumn,
  superscriptaddress,
  longbibliography,
  nofootinbib
]{revtex4-2}
\usepackage{dcolumn}
\usepackage{bm}
\usepackage{hyperref}
\usepackage{amsthm}
\usepackage{graphicx}
\usepackage[makeroom]{cancel}
\usepackage{comment}
\hypersetup{
    colorlinks = true,
    allcolors=black,
    citecolor=blue,
    linkcolor=blue
}
\usepackage{dsfont, mathrsfs, float}
\usepackage[caption=false]{subfig}
\usepackage{orcidlink}
\usepackage{xcolor}
\usepackage{mathtools}
\usepackage{multirow}
\usepackage[capitalise]{cleveref}

\newcommand{\m}[1]{\mathcal{#1}}
\newcommand{\mr}[1]{\mathrm{#1}}

\begin{document}
\title{\textcolor{black}{Storage, Scrambling, and Loss of Information in Quantum Reservoir Computing}}
\author{Nathan Keenan}
\affiliation{Instituto de Física Interdisciplinar y Sistemas Complejos (IFISC), UIB–CSIC
UIB Campus, Palma de Mallorca, E-07122, Spain}
\author{Roberta Zambrini}
\affiliation{Instituto de Física Interdisciplinar y Sistemas Complejos (IFISC), UIB–CSIC
UIB Campus, Palma de Mallorca, E-07122, Spain}

\begin{abstract}
\textcolor{black}{The performance of a quantum reservoir computer in temporal processing tasks depends on how its driven quantum substrate retains information about past inputs, distributes it across physical degrees of freedom, and loses it through environmental dissipation and measurement feedback. We formulate these processes using a classical-quantum state obtained by restricting a reservoir process tensor to classical input encoding and single-time readout. Conditional subsystem Holevo quantities describe information about selected input histories and bound its accessibility to measurements on subsystems of the reservoir. In a six-qubit all-to-all transverse-field Ising reservoir, we find that the total stored information changes relatively little across Hamiltonian parameters, while its spatial distribution and temporal decay vary strongly. Effective diagnostics of these two behaviours identify different regions of high information-processing capacity for linear and higher-degree temporal tasks. Measurement-induced dephasing can improve noiseless task performance when it increases forgetting rates without strongly reducing information delocalisation. The framework separates storage from subsystem accessibility and provides a common description of information flow in driven quantum learning systems.}
\end{abstract}
\date{\today}
\maketitle

\section{Introduction}
Reservoir computing (RC) \cite{nakajima2020physical, nakajima2021reservoir} offers a computational platform specialised in the processing of temporal data. Time series data is sequentially loaded into a computational substrate, which acts as a bath of complexity with memory: previously injected information is mixed and spread across many degrees of freedom of the substrate. Alongside the data that is injected, data are sequentially read out from the dynamical system, and passed through a linear readout layer. This linear layer can be trained in order to perform a variety of target tasks: short term memory recall, time series forecasting, and other linear and non-linear functions across the temporal input data. 

Quantum reservoir computing (QRC) \cite{Fujii2017harnessing,nokkala,mujal2021opportunities} platforms are those whose dynamical substrate is a quantum system. The exponential scaling of Hilbert space dimension with the number of physical units is appealing for the separation in state space of evolutions arising from different input histories, which should give rise to better trainability of the readout layer. Moreover, one can also consider direct processing of quantum temporal data without having to perform a classical encoding. There are some subtleties when moving to a quantum substrate, however. For one, performing measurements on the reservoir to pass data onto the output layer is no longer a passive activity due to measurement backaction on the substrate \cite{mujal2023time,oriol2026}. Moreover, the large Hilbert space dimension is not simply a benefit to RC platform performance, and the curse of dimensionality can arise through concentration of observable expectation values around thermal values \cite{xiong2025role, sannia2025exponential}.

While QRC performance is usually characterized by benchmark tasks \cite{dambre2012information}, we still lack an information-theoretic description of how temporal information is stored, distributed and eventually lost inside the quantum substrate. Furthermore, existing analyses of quantum reservoir computing focus primarily on the instantaneous state of the reservoir. However, reservoir computing is inherently a temporal information-processing task: even when the injection and read-out are instantaneous, the relevant object is not a single-time quantum state, but the relationship between a history of inputs and the evolving reservoir state. At present, although there are some recent works analysing some aspects of the temporal structure of information in quantum reservoir computing platforms \cite{kobayashi2025quantum, ding2026thermodynamics, wang2026fisher}, there is still no compact information-theoretic description of the full temporal input-output structure.

\begin{figure*}
        \includegraphics[width=\linewidth]{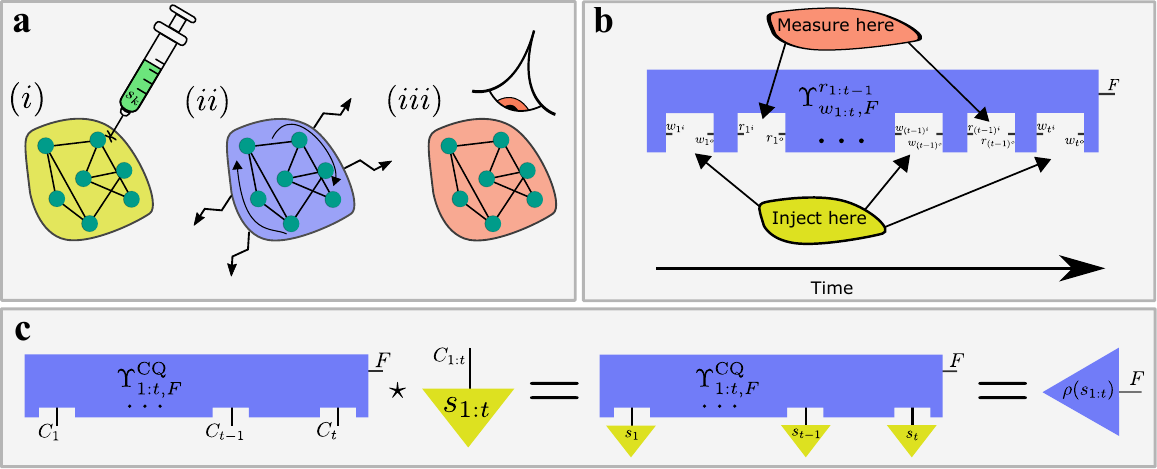}
    \caption{\textbf{a}: Schematic illustrating the quantum reservoir computing procedure. (i) Information (classical or quantum) is injected into the quantum substrate. (ii) The substrate then undergoes (unitary or non-unitary) evolution. Information about previously injected inputs becomes spread across the degrees of freedom in the system, and perhaps lost to the environment. (iii) Measurements are made on the substrate in order to extract the `processed' information, which is then sent to a linear readout layer (not pictured). The entire procedure is repeated for the next input. Note that this exact pattern may be altered, for example in the use of temporally multiplexed measurements. \textbf{b}: Process tensor for the QRC protocol. The tensor $\Upsilon_{w_{1:t}, F}^{r_{1:t-1}}$ has slots for interventions corresponding to data injection, and slots for interventions corresponding to data extraction, labelled $w$ for \textit{write} and $r$ for \textit{read} respectively. The positions of the slots do not indicate the portion of the system they are acting on. For example, reading and writing may be carried out on different parts of the system. The space $F$ corresponds to the output state after step $t$ of the QRC procedure. \textbf{c}: After simplifying the reservoir process tensor for commonly used QRC setups, one gets the CQ state $\Upsilon^{\mr{CQ}}_{1:t, F}$ that can be contracted with a choice of inputs $s_{1:t}$ to yield the resulting reservoir state $\rho(s_{1:t})$.}
    \label{fig:first}
\end{figure*}

In this work, we show that the process tensor framework \cite{milz2017introduction, milz2021quantum, taranto2025higher} is a natural language for describing the dynamics of the QRC process, which is illustrated in \cref{fig:first} \textbf{a}. Under the assumptions commonly used in QRC, the representative process tensor shown in \cref{fig:first} \textbf{b} reduces to a classical-quantum (CQ) state that represents the dependence of the reservoir state on the complete input history.

The CQ state represents the inject-evolve-measure-repeat process of the QRC protocol and is illustrated in \cref{fig:first} \textbf{c}. Mutual informations between the classical and quantum subsystems of this state are given by Holevo quantities \cite{holevo1973bounds, schumacher1997sending, holevo1998capacity, holevo1998coding, nielsen2010quantum} which can be used to study memory storage in the system of past classical inputs. In particular, we use these Holevo quantities to study spatial distribution of injected information within a well-known QRC platform, along with the temporal decay of information of historical inputs. We broadly categorise the dynamics of the information injected into the substrate into two behaviours: scrambling of information within the system whereby initially local information spreads into nonlocal degrees of freedom, and loss of information from the system due to dissipative dynamics. These behaviours are both necessary for the proper functioning of a QRC platform \cite{martinez2021dynamical, kobayashi2026edge, sannia2024dissipation, kobayashi2024extending, martinez2023quantum}, but have also been found to have a complex relationship between each other for general open quantum systems \cite{zanardi2021information, zhang2023dynamical}. We characterise these two behaviours through the aforementioned Holevo quantities, and display their empirical relationship to the performance of our QRC model by analysing the information processing capacity (IPC) \cite{dambre2012information}.

This work is structured as follows. First, we give some background context to the current QRC program in \cref{sec:background}. We do this in order to make explicit the framework we are using, which corresponds to the commonly used QRC protocol 
where classical data is injected into the system, and only single-time expectation values are passed to the linear readout layer.

In \cref{sec:holevo} we then introduce the CQ state $\Upsilon^\mathrm{CQ}_{1:t, F}$, which has as Hilbert space $\m H_F$ not only for the output state of the QRC protocol after step $t$, but also classical auxiliary Hilbert spaces $\m H_{C_k}$ corresponding to each input $s_k$. We believe the approach followed here for studying QRC platforms makes apparent the link between data injection and extraction from the QRC substrate,  and the wider literature of intervention-based studies of many-body quantum systems \cite{dowling2024operational, o2026diagnosing}. Therefore in \cref{appendix:framework} we provide the derivation of this CQ state from the process tensor describing the QRC protocol as seen in \cref{fig:first} \textbf{b}. Back in the main text, we then introduce the Holevo quantities that represent quantum mutual informations between different subspaces of the output space $F$ and subsets of the classical input spaces $C_1,\cdots C_t$. Such Holevo quantities are commonly used to study information capacity in quantum systems, and have even recently been used to measure predictive capacities of QRC platforms \cite{ding2026thermodynamics}.

In \cref{sec:numerics} we use the tools explicated in the previous sections to study the case of local injections in a commonly used QRC substrate: the disordered all-to-all transverse-field Ising model. By tuning the average strength and disorder of the transverse field in this model, we are able to sweep dynamical regimes with different information processing capabilities, which we diagnose via the scrambling and loss of information.
We then use the IPC to investigate the performance of this QRC setup across the different dynamical regimes, and discuss the implications of these results in light of the so-called memory-nonlinearity trade-off \cite{dambre2012information}.

Lastly in \cref{sec:discussion}, we suggest possible future directions of this work. In particular, we comment on how the framework used here can be expanded in the future to go beyond the single-time-expectation-value framework, and hopefully take better advantage of the quantum properties of the substrate for processing temporal data.

\section{QRC Framework}
\label{sec:background}
The QRC platform can be broadly modularised into four design elements: (a) a choice of substrate for the quantum reservoir (which may have unitary or non-unitary dynamics) (b) a method of injecting data into the system, (c) a measurement setup for extracting data from the reservoir, and (d) a loss function for training the output layer for a given task. Putting aside (d) for the moment, the current QRC literature contains a wide variety of ways of dealing with modules (a)-(c). Representative substrates include qubit and spin systems spanning superconducting quantum annealers, neutral-atom processors, and nuclear spins in molecular solids\cite{Fujii2017harnessing,nakajima2019boosting,kutvonen2020optimizing,martinez2021dynamical,mujal2023time,sannia2024dissipation,chen2020temporal,Sakurai2022,hou,domingo,gotting,
sannia2026temporal,gyurik2026quantum,kornjavca2024large,negoro2018machine}, bosonic systems, including continuous- and discrete-variable photonic platforms, Bose--Hubbard lattices, and nonlinear oscillators \cite{paparelle2026experimental,nokkala,garcia2025quantum,spagnolo2022experimental,di2026time,sakurai2025quantum,llodra2025quantum,govia2021quantum}; fermionic systems \cite{ghosh2019quantum,llodra2023benchmarking,kobayashi2026edge}; and hybrid spin-boson systems \cite{Senanian2024,carles2026experimental,das2025quantum}. These examples illustrate the diversity of substrates rather than provide an exhaustive classification.

For classical data, examples of input encoding include the Fujii--Nakajima erase-and-write map \cite{Fujii2017harnessing,nokkala}, input-dependent unitary operations or Hamiltonian control, as used in circuit-based and analog reservoirs \cite{li2026quantum,sannia2024dissipation,chen2020temporal,kornjavca2024large}, coupling the reservoir to incoming optical modes carrying the encoded data \cite{garcia2023scalable}, and modulation of dissipative rates \cite{Sannia_skin}. Quantum inputs have also been considered in quantum reservoir processing \cite{ghosh2019quantum} and quantum extreme learning machines \cite{wang2022review,gili2026learning}, as well as in temporal processing with hybrid quantum-classical reservoirs \cite{nokkala2024retrieving,coll2026temporal}.

For the data extraction procedure, a POVM can be implemented across the QRC protocol \cite{mujal2023time}. The choice of POVM is important, as it defines which modes of the system are being read out \cite{oriol2026}. Although early works in the area ignored measurement backaction on the substrate, measurement-induced decoherence can also be utilised in order to assist with ensuring the fading memory property of the platform \cite{mujal2023time,sannia2024dissipation} or even as an input-injection mechanism \cite{Sannia_skin,Hu2024,oriol2026}.

The trainable output layer is linear and time-local, and so the bulk of the complexity of the QRC platform is contained in the previous three components. Non-linearities in the input are generally encoded in the injection procedure, the dynamics of the reservoir scrambles these inputs across local and nonlocal modes in the system, and a choice of POVM measurement retrieves a particular portion of the scrambled information of the inputs. After the training of the linear layer by minimising a loss function across the measurement training data, one can then calculate various performance metrics across a test batch of data. Some common targets involve short-term memory tests -- how well can the readout layer retrieve past inputs to the reservoir -- along with the family of more general non-linear memory targets that together with the former make up the information processing capacity (IPC) \cite{dambre2012information, martinez2023information}.

\subsection{Dynamics of the QRC protocol}
\label{sec:setup}
Before we define the process tensor for the QRC protocol, we will first illustrate in more detail the \textit{inject-evolve-measure} cycle performed on the reservoir substrate. This cycle provides a common operational description of many QRC implementations, while making explicit the ordering of input injection, reservoir evolution, and measurement adopted in this work.

In order to process the time series $\{s_{1:t}\}$ (of either classical or quantum data),\footnote{For classical data, $s_k$ is usually a real number, while for quantum data $s_k$ is usually a quantum state.} the \textit{inject-evolve-measure} procedure illustrated in \cref{fig:first} is given by:
\begin{enumerate}
    \item At step $k$, inject data point $s_k$ into the system
    \item Allow the reservoir to evolve with its native dynamics via the dynamical map $\m E_k$
    \item Perform measurements on the system via the single-time instrument $\m J = \{\m M^{(x)}\}$
    \item Repeat all steps above for step $k+1$.
\end{enumerate}
This method of \textit{sharp} injections is carried out by applying a CPTP map parameterised by the input data on all or part of the reservoir substrate. One could also imagine injecting data by varying some parameters in the dynamical map of the substrate \cite{sannia2024dissipation, ding2026thermodynamics}, however we do not comment on this \textit{continuous} injection method in this work.

In order to increase the amount of output data, one may also make use of time multiplexing \cite{vcindrak2025krylov}, which involves repeating steps (2) and (3) above many times before injecting the next data point. Of course there are many other possible sequences of operations using the actions inject, evolve, measure, that may be physically or operationally motivated. However, in this work we will stick with the above standard realisation. 

One important thing to note that makes this framework amenable to our CQ state representation is that only single-time expectation values from this process are passed to the linear readout layer. This means that all data stored in multi-time correlations in the reservoir substrate go unused, and the information that the readout layer receives at step $t$ is contained within the output state of the substrate at step $t$. As a result, for each past measurement performed on the substrate with the instrument $\m J = \{\m M^{(x)}\}$, the average measurement map $\tilde{\m M} = \sum_x \m M^{(x)}$ is applied on the state of the reservoir. We discuss potential future extensions of the QRC program in \cref{sec:discussion} in the context of passing instead multi-time data to the readout layer, but for this work we stick with the restricted single-time setup.

\subsection{Training the Output Layer}
\label{sec:train}
Usually the output data across the whole QRC experiment is divided into three groups based on the cycle number. The first $n_\mathrm{washout}$ timesteps are discarded: one of the necessary properties for a QRC to process time series data is to have a fading memory of the initial state, and one usually discards the output data while waiting for the reservoir to forget its initial conditions. 

The following $n_{\mr{train}}$ cycles of data are used to train the output layer given a target task. Let $ y_k = f_k(s_{1:t})\in \mathbb{R}$ be some target function of the time series data $s_{1:t}=(s_1,\cdots,s_t)$. For short term memory retrieval, one has $f_k( s_{1:t}) = s_{k-\tau}$, and for time series forecasting, one has 
\begin{equation}
    f_k(s_{1:t}) = s_{k+\tau}.
\end{equation}
For the case of training with single-time expectation values, we have a vector $\mathbf x_k\in \mathbb R^M$ of measurement data from step $k$, and the goal is to train an output layer $\mathbf w\in \mathbb{R}^{M}$ in order to minimise some cost objective function between the measurement feature data and target data:
\begin{equation}
    \mathbf{w}^* = \mr{argmin}_{\mathbf{w}}\left(\sum_{k\in \mr{training}}\m L(\mathbf{w}\cdot \mathbf{x}_k,  y_k)\right),
\end{equation}
which might also include a regularisation factor in $\vec w$ to prevent overfitting. This trained output layer is then used on the output data from the remaining steps in order to (hopefully) well-estimate the target function on those timesteps. In particular, it will be relevant for \cref{sec:numerics} when we calculate reservoir IPCs. 

\subsection{Process tensors and the QRC protocol}
The inject-evolve-measure cycle described above is a particular instance of a more general problem: characterising how a sequence of interventions on a quantum system is related to the resulting statistics of that system. This is precisely the setting addressed by the process tensor formalism \cite{milz2017introduction,milz2021quantum,taranto2025higher}, which was originally developed to give a description of open quantum systems subject to repeated interventions. In this language, a QRC platform can be viewed as a fixed background process (the reservoir dynamics) that is repeatedly probed by external interventions -- the injection of classical data and the extraction of measurement outcomes -- with the process tensor itself encoding all correlations between these interventions and the resulting output states, including any memory effects mediated by the reservoir. This perspective is natural for QRC precisely because reservoir computing is a fundamentally temporal task: the object of interest is not the reservoir state at a single time, but the map from an entire history of inputs to the evolving output state. Framing the QRC protocol in this way also makes explicit its connection to the broader literature on intervention-based characterisations of many-body quantum systems, where process tensors have been used to probe information scrambling and related dynamical phenomena. In \cref{appendix:framework} we make this connection concrete: starting from the process tensor $\Upsilon^{r_{1:t-1}}_{w_{1:t},F}$ associated with the general QRC protocol of \cref{fig:first} \textbf{b}, we show how the assumptions of deterministic classical injections and single-time readout reduce this object to the classical-quantum state $\Upsilon^{\mathrm{CQ}}_{1:t,F}$ introduced in \cref{sec:holevo}, thereby providing the technical derivation underlying the Holevo quantity framework used throughout the remainder of this work.

\section{The CQ state and Holevo quantities}
\label{sec:holevo}
In \cref{appendix:framework} we start with a general process tensor $\Upsilon_{w_{1:t}, F}^{r_{1:t-1}}$ representing the injection and extraction of information during a QRC protocol up to some time $t$ illustrated in \cref{fig:first} \textbf{b}, and show that in the case of classical deterministic injections and the passing of only single-time expectation values to the readout layer, this process tensor reduces to the classical-quantum (cq) state:
\begin{equation}
\label{eq:cq}
    \Upsilon_{1:t, F}^{\mr{CQ}}= \int_{\Omega^t}\rho(s_{1:t})\otimes |s_{1:t}\rangle\langle s_{1:t}| \ \mathrm{d}s_{1:t},
\end{equation}
where for this work we choose the compact domain of classical inputs to the reservoir to be $\Omega := [0,1]$.

Note that contracting a choice of input history $|s_{1:t}\rangle\langle s_{1:t}|$ with the CQ state $\Upsilon_{1:t, F}^{\mathrm{CQ}}$ yields the corresponding output state $\rho(s_{1:t})$ of the substrate, and this is illustrated in \cref{fig:first} \textbf{c}. 

For sharp injections via CPTP maps $\mathcal A^{(s)}$, average measurement map $\tilde {\mathcal M}$, and dynamical map $\mathcal E$ in between steps, the output state given history $s_{1:t}$ is given inductively by
\begin{equation}
\label{eq:update}
    \rho(s_{1:t}) = (\m E \circ \mathcal A^{(s_t)}\circ \tilde{\m M}) [\rho(s_{1:t-1})].
\end{equation}

The Holevo quantity\footnote{Note that as a Holevo quantity, $\chi_t$ implicitly assumes an independent, uniform sampling of the $s_k$. This does not mean that only uniform independent data can be applied to the classical sites of the process tensor; it is merely an artifact of our convention for instantiating the deterministic interventions in the process tensor, and $\chi_t$ serves merely as a diagnostic tool. A different convention may be chosen to target a specific input distribution, although the diagnostic benefits of doing this are not so clear.} \cite{holevo1973bounds, schumacher1997sending, holevo1998capacity, holevo1998coding, nielsen2010quantum} appears directly as the quantum mutual information (QMI) of the bipartite CQ state $\Upsilon^{\mathrm{CQ}}_{1:t,F}$ between the classical and physical subsystems:
\begin{align}
    \chi_t &= I(C_{1:t} ; F) =  S(\Upsilon_{{1:t}}^C) + S(\Upsilon_F^Q) - S(\Upsilon^{\mathrm{CQ}}_{1:t,F})\nonumber \\ &=S\left(\int_{\Omega^t}\rho(s_{1:t}) \ \mathrm{d}s_{1:t}\right) - \int_{\Omega^t} S(\rho(s_{1:t})) \ \mathrm{d}s_{1:t},
\end{align}
where $\Upsilon_F^{Q}$ and $\Upsilon_{1:t}^\mathrm{C}$ are obtained from \eqref{eq:cq} via partial trace, $S(\rho) = -\mr{Tr}(\rho\log\rho)$ is the von-Neumann entropy, and $C_{1:t}=C_1\otimes \cdots  \otimes C_t$.

The Holevo quantity has indeed a useful physical interpretation: it upper bounds the amount of information that can be retrieved about past inputs $s_{1:t}$ per shot of POVM measurement on the output state $\rho(s_{1:t})$.
This object turns out to be quite useful for addressing ideas such as fading memory and nonlocal scrambling of information, the latter of which can be probed by calculating Holevo quantities on subsystems of both the past intervention sites, and the final physical space $F$. 

Since $\chi_t$ upper bounds the amount of information per-shot that any POVM can extract about the inputs $s_{1:t}$, it might be tempting to use the non-zero value and/or growth of $\chi_t$ as a function of $t$ to guarantee the required separation property of the QRC platform. However, these are not strong enough requirements as the choice of POVM used to extract information may significantly underperform compared to the upper bound found in $\chi_t$. Nonetheless, the behaviour of $\chi_t$ as a function of $t$ is interesting from the perspective of information saturation of the substrate, and this will be numerically studied in \cref{sec:numerics}. In the following, we will introduce conditional and local Holevo quantities that can be used to assess the spatio-temporal behaviour of information across the system. These can be related to the fading memory and information scrambling in QRC.
\subsection{Fading memory \& conditional Holevo quantities}
\label{sec:fading}
\begin{figure}
    \centering
        \includegraphics[width=0.85\linewidth]{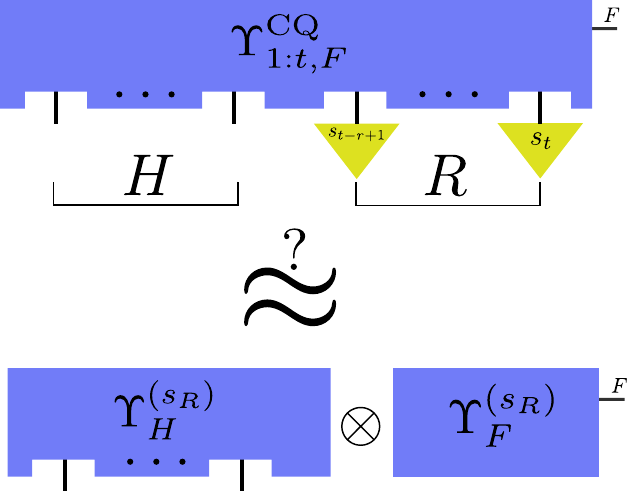}
    \caption{The CQ state $\Upsilon^{\mr{CQ}}_{1:t, F}$ with input spaces split into `recent' ($R$) and `historical' ($H$). The inputs of the $R$ window are then contracted with a choice of the classical variables, leaving only the $H$ legs open. Lower memory in the final state of historical inputs from $H$ will lead to better approximation of this partially-contracted CQ state by a product structure.}
    \label{fig:markov}
\end{figure}

Recall that $\chi_t = I(C_{1:t}; F)$ is the QMI between the classical input spaces and the output space of the quantum state after $t$ steps. We can instead ask about the conditional QMI between a \textit{subset} of the classical input spaces and the final output space. Of particular interest will be the QMI between the final output space and the inputs from historical steps $H=1:t-r$,  conditioned on the recent inputs from steps $R=t-r+1:t$. This will allow us to ask about the memory the output state has about inputs that occurred at least $r$ steps ago. This conditional QMI is calculated between $H$ and $F$ in the process tensor illustrated in \cref{fig:markov}, and can be written as the full conditional Holevo quantity:
\begin{equation}
\label{eq:holcond}
    I(C_H ; F |C_{R}) = \int_{\Omega^r}\chi_t|_{s_{R}} \ \mathrm{d}s_{R} =:\chi_{t}{(r)} ,
\end{equation}
where $\chi_t|_{s_{R}}$ is a pointwise-conditional Holevo quantity for a given recent input sequence $s_R$:
\begin{align}
\label{eq:marginal}
    \chi_t|_{s_{R}}= S&\left(\int_{\Omega^{t-r}}\rho(s_H, s_R) \ \mathrm{d}s_{H}\right)\nonumber \\ &- \int_{\Omega^{t-r}} S(\rho(s_H, s_R)) \ \mathrm{d}s_{H}.
\end{align}
where $\rho(s_H, s_R)\equiv\rho(s_{1:t})$.

Another way to think of the pointwise-conditional Holevo quantity, is given a `tester' $|s_R\rangle\langle s_R|$ over recent input window $R$, $\chi_t|_{s_{R}}$  defined in \cref{eq:marginal} can be calculated as a quantum relative entropy between (a) the full process tensor and (b) its approximation as a product between reduced states on output space and historical inputs:
\begin{equation}
    \chi_t|_{s_{R}} = D(\Upsilon_{HF}^{(s_R)} || \Upsilon_{H}^{(s_R)} \otimes \Upsilon_{F}^{(s_R)}),
\end{equation}
where 
\begin{align}
    &\Upsilon_{HF}^{(s_R)} = \Upsilon_{1:t, F}^{\mr{CQ}}\star |s_R\rangle\langle s_R|, \nonumber\\ &\Upsilon_{H}^{(s_R)} = \mr{tr}_F\Upsilon_{HF}^{(s_R)}, \nonumber\\ &\Upsilon_{F}^{(s_R)} = \mathrm{tr}_{{H}}\Upsilon_{HF}^{(s_R)}.
\end{align}
Vanishing values of $\chi_t|_{s_{R}}$ in increasing `recent window' width $r$ thus imply an approach to a product structure in \cref{fig:markov}, 
and the full conditional Holevo quantity $\chi_t(r)$ in \cref{eq:holcond} can be understood as an average deviation from product structure of the temporally-correlated state displayed in \cref{fig:markov}.

In order to link the decay of the above conditional Holevo quantities on works that use the trace distance in studying the fading memory/echo-state requirement \cite{martinez2023quantum, kobayashi2024coherence, kobayashi2024extending} a Pinsker-type inequality may be used \cite{kullback1967lower}.

\subsection{Information scrambling \& subsystem Holevo quantities}
Holevo quantities have been used previously to analyse information scrambling in quantum systems \cite{zhuang2022phase, nico2022memory, yuan2022quantum, zhuang2023dynamical, chen2026probing} by probing how much information that was initially stored locally in a system has spread nonlocally during the native evolution of the substrate. In order to study the accessibility of past information to local probes we can use QMIs restricted to subsystems of the full final Hilbert space. For example, given the bipartition $F=F_1\otimes F_2$ of the final state space, the total correlations between subsystem $F_1$ and historical input spaces $H$ is given by 
\begin{equation}
    I(C_{H} ; F_1 |C_{R}) = \int_{\Omega^r}\chi_t|_{s_R,F_1} \ \mathrm{d}s_{R} =:\chi_{t}(r)|_{F_1} ,    
\end{equation}
where
\begin{align}
\chi_{t}|_{s_R, F_1} = S&\left(\int_{\Omega^{t-r}}\mr{tr}_{F_2}\rho(s_H, s_R) \ \mathrm{d}s_{H}\right) \nonumber \\ &- \int_{\Omega^{t-r}} S(\mr{tr}_{F_2}\rho(s_H, s_R)) \ \mathrm{d}s_H.
\end{align}
We will then use the notation $\chi_{t}(r,f)$ for the average of $\chi_{t}(r)|_{F_1}$ over all choices of $F_1$ consisting of $f$ physical subunits (qubits, in our case)\footnote{Notice that because the reservoir has an all-to-all connectivity, \textcolor{black}{apart from inclusion/exclusion of the writeable qubit}, subsystems of the same size are statistically equivalent after averaging over disorder realizations. Averaging the Holevo quantity over all subsets of size $f$ therefore provides a meaningful characterization of the spatial distribution of information.}:
\begin{equation}
    \chi_t(r,f) = \frac{1}{{N \choose f}}\sum_{|F_1| = f}\chi_{t}(r)|_{F_1}.
\end{equation}
Whereas $\chi_{t}(r)$ measures the memory that the final state has of historical inputs, $\chi_{t}(r,f)$ measures how much of that memory on average is stored in subsystems of size $f$. Therefore we can analyse and compare the behaviours of $\chi_{t}(r,f)$ for different values of $f$ in order to try and separate the loss of information due to the dissipative nature of the injection protocol/dynamical map/measurement backaction, from the local loss of information due to scrambling across the reservoir.

\section{Results}
\label{sec:numerics}

We now apply the Holevo-quantity formalism introduced above to a representative QRC platform, providing a spatiotemporal characterisation of its information-processing capabilities and illustrating the diagnostic power of this approach.

\subsection{The QRC Platform}

Let us identify the injection, evolution, and measurement steps of the QRC dynamics defined in \cref{eq:update}.

\begin{figure}
    \centering
        \includegraphics[width=\linewidth]{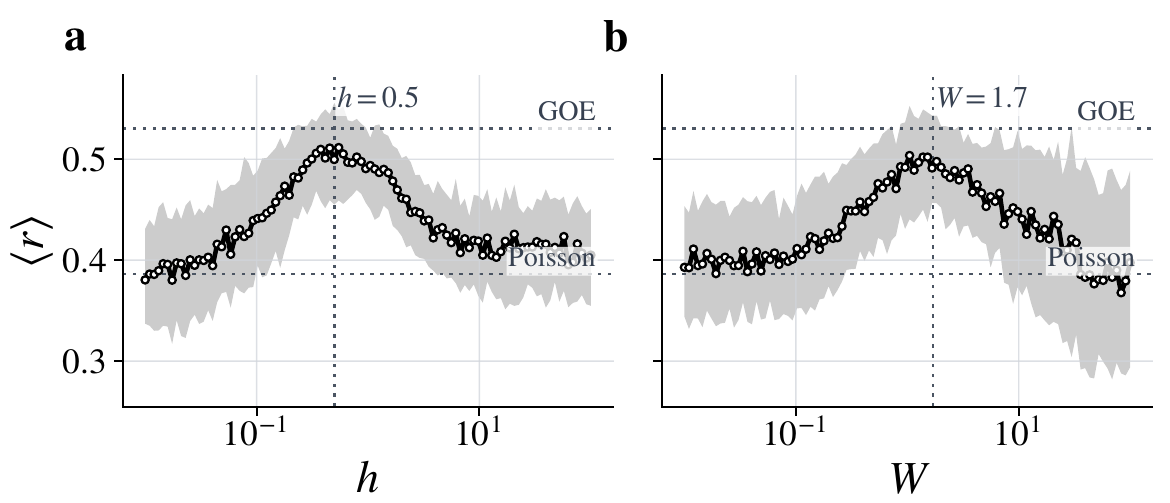}
    \caption{Level spacing statistics for the two main parameter sweeps investigated in this section. \textbf{a}: The first sweep is along $h$ values for $W=0$. \textbf{b}: The second sweep is along $W$ values for $h=0.03$. In both cases, we indicate the values of the mean level spacing ratios from Poissonian (integrable) and GOE (chaotic) statistics, along with vertical markers to indicate the parameter values at the `maximally chaotic' points in each cut. The markers indicate ensemble median across Hamiltonian samples, with a faint window indicating the $16-84$ percentile range of results. The full phase diagram is represented in \cref{fig:phase} of the Appendix.}
    \label{fig:cuts}
\end{figure}

\subsubsection{The Hamiltonian}
We choose the disordered all-to-all transverse-field Ising model as our QRC substrate because its rich phase diagram, spanning integrable, chaotic, and many-body-localised regimes as the field strength $h$ and disorder $W$ are tuned, has been shown in several independent studies to strongly affect QRC performance~\cite{martinez2021dynamical, palacios2024role, Xia2021}, making it an ideal test-bed for linking our information-theoretic diagnostics to the underlying dynamical regime. The substrate Hamiltonian reads:
\begin{equation}
\label{eq:hamil}
    H = \sum_{j<k}J_{jk} X_j X_k + \sum_j (h+w_j)Z_j,
\end{equation}
where $J_{jk} \sim U([0, 1])$, and $w_j\sim U([-W, W])$. 

We will analyse results along two specific parameter sweeps. The first sweep varies the average transverse field strength $h$ at vanishing disorder $W=0$, while the second varies the disorder strength $W$ at fixed average field $h=0.03$. We plot the mean level spacing ratios of these regimes in \cref{fig:cuts}. We also include a phenomenological phase diagram in $(h, W)$ of the model in \cref{appendix:numerical}. Along the first sweep, increasing $h$ drives the model to a chaotic transition reached around $h\approx0.5$ for the $6$-site model. Along the second sweep, by increasing W there is a chaotic transition as before around $W\approx1.7$, but this time alongside a disorder-induced transition from chaotic to localised behaviour for large $W$.

\subsubsection{Injections \& Measurements}
For the injection of data into the substrate, we will use the erase-and-write map:
\begin{equation}
\label{eq:eaw}
    \m A^{(s)}[\rho] = \sigma^{(s)} \otimes \mathrm{tr}_1(\rho),
\end{equation}
that traces out the first qubit and replaces it with the state $\sigma^{(s)}$, which has the following form:
\begin{align}
    \sigma^{(s)} &= |\psi^{(s)}\rangle\langle\psi^{(s)}|, \label{eq:eawt} \\ |\psi^{(s)}\rangle &= \sqrt{1-s}|0\rangle + \sqrt{s}|1\rangle \label{eq:eawtt}.
\end{align}

Accounting for measurements throughout the protocol is necessary to correctly capture their back-action on the reservoir dynamics, and because online monitoring schemes require measurement to be treated as an integral part of the processing itself, rather than a final readout step~\cite{mujal2023time,oriol2026}.
Weak measurements will be carried out across the substrate in the $Z$ basis. For the calculation of Holevo quantities, this choice of measurement is only used in order to specify what the average measurement map $\tilde{\mathcal{M}}$ will be across the protocol. The measurement strength $g$ parameterises the average measurement map $\tilde{\mathcal M}_j$ of $Z_j$ in the following way \cite{mujal2023time}
\begin{equation}
\label{eq:measure}
    \tilde{\mathcal M}_j[\rho] = \left(\frac{1+e^{-g^2/2}}{2}\right)\rho +\frac{1-e^{-g^2/2}}{2}Z_j\rho Z_j,
\end{equation}
where $\tilde{\mathcal M} = \prod_j \tilde{\mathcal M}_j$. In order to analyse the effect of measurement separately, we assume vanishing measurement strength in the main analysis, and study the effects of finite measurement strength in \cref{sec:measurement}. Further details for simulation parameters can be found in \cref{appendix:params}.

\subsection{Holevo quantities \& memory behaviour}
\begin{figure}
    \centering
        \includegraphics[width=\linewidth]{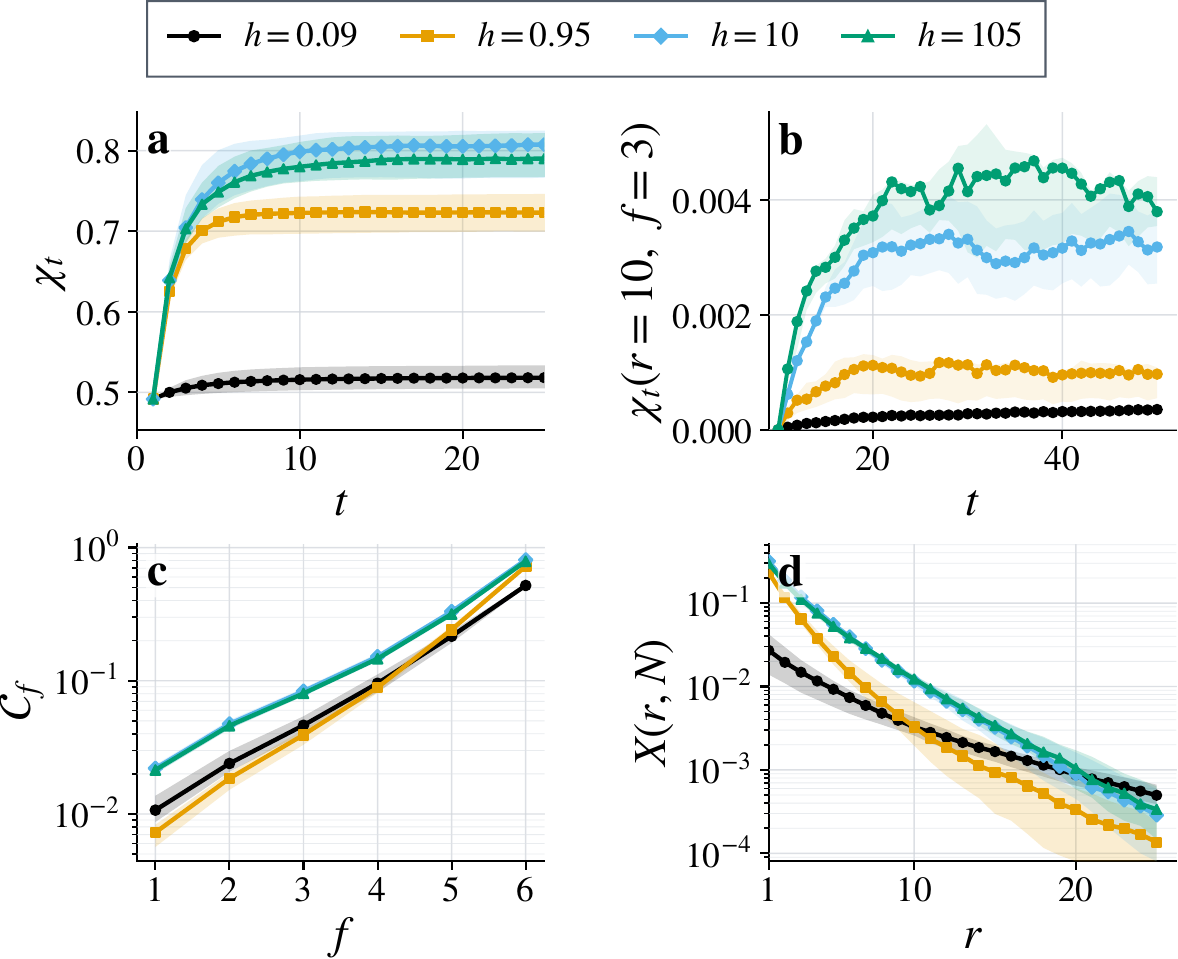}
    \caption{\textbf{a}: The full Holevo value as a function of number of inputs $t$.
    \textbf{b}: Conditional Holevo quantities for $r=10$ and $f=3$ as a function of $t$.
    \textbf{c}: Stationary subsystem capacities $\m C_f$ as a function of system size $f$, probed using numerics at $t=50$. 
    \textbf{d}: Full system conditional Holevo value $X(r, N)$ as a function of conditioned window width $r$, probed using numerics at $t=50$ steps of the process. All panels show results for Hamiltonians with $W=0$ and a selection of different $h$ values. The markers indicate ensemble median across Hamiltonian samples, with a faint window indicating the $16-84$ percentile range of results. The results are gathered after $1000$ washout steps.}
    \label{fig:results0}
\end{figure}

Let us now apply the tools we introduced in \cref{sec:holevo} to characterize the spatio-temporal information evolution in the Ising QRC.
The Holevo information across all past inputs grows with the input length $t$, as seen in \cref{fig:results0} \textbf{a}, reflecting the build-up of distinguishable states across input histories, up to a saturation value. While a longer input history can in principle make the states more distinguishable, monotonicity is not guaranteed in general, since it also depends on the effect of the map updating the state at each step. A saturation value is expected on general grounds (bounded by $\log\dim\mathcal{H}_F$), but whether and how it is reached depends on the specific choice of dynamical map. In our case we see that this saturation value lies about an order of magnitude lower than the theoretical maximum across $h$ and $W$ values. Our results indicate that at $t=50$,  we are able to probe stationary values for $\chi_t(r,f)$ for the values of $r$ and $f$ studied here, as seen for the Holevo quantities displayed in \cref{fig:results0} \textbf{a}-\textbf{b}. We use $X(r,f)$ to indicate these stationary values, and we define the empirical \textcolor{black}{Holevo} capacity $\mathcal{C}_f :=X(0, f)$ of subsystems of size $f$ in our QRC setup.

We now turn our attention to studying the growth of \textcolor{black}{Holevo capacity} $\m C_f$ in subsystem size $f$, along with the decay of memory $X(r,f)$ in $r$ -- shown in \cref{fig:results0} \textbf{c} \& \textbf{d} respectively.

\subsubsection{Full- \& Sub-system memory capacities}
\begin{figure}
    \centering
        \includegraphics[width=\linewidth]{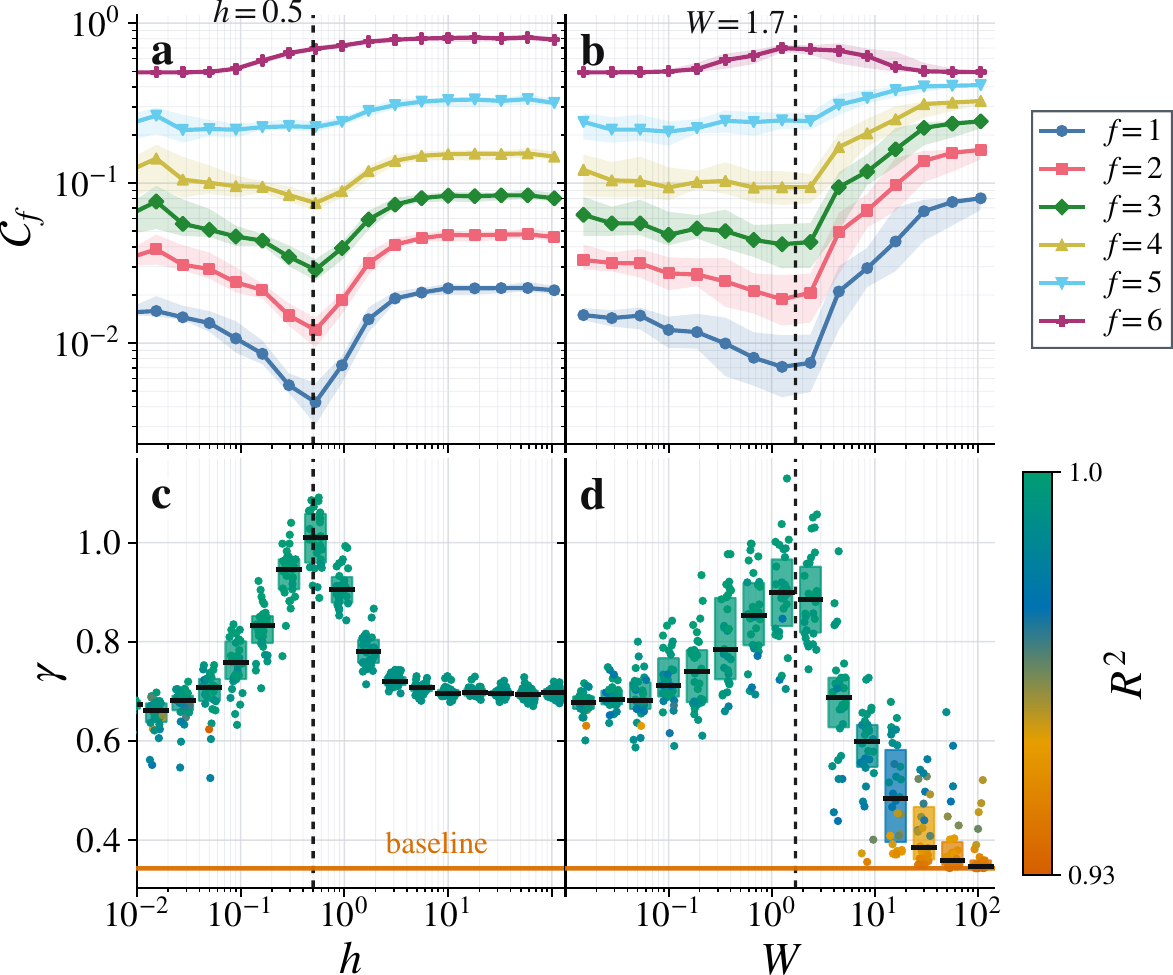}
    \caption{\textbf{a}-\textbf{b}: The input-specific memory capacity $\mathcal{C}_f$ of the subsystems of size $f$ as a function of Hamiltonian parameters $h$ and $W$. \textbf{c}-\textbf{d}: Extracted growth rate $\gamma$ of information storage across subsystem sizes $f$. Results across $h$ values are along the cut $W=0$, while results across $W$ values are along the cut $h=0.03$. The extracted growth rates are shown on a scatter plot across Hamiltonian samples, colour-coded by the $R^2$ value of the fit. Horizontal black lines indicate median values, while the extended rectangles display a $16$-$84$ percentile range of results. These choices are in order to display bulk behaviour while avoiding obscuring effects of outliers. \textcolor{black}{The horizontal line labelled `baseline' in \textbf{c}–\textbf{d} indicate $\gamma_{\mathrm{base}}\simeq0.343$ for the non-interacting reference $C_f=(f/N)C_N$, with input information confined to the injection qubit. Their colour indicates the corresponding fit quality, $R^2\simeq0.936$.}}
    \label{fig:results1}
\end{figure}
In \cref{fig:results1} \textbf{a}-\textbf{b} we plot the subsystem capacities $\mathcal{C}_f$ as a function of Hamiltonian parameters across different subsystem sizes $f$. The memory capacity increases systematically with subsystem size as expected, while its dependence on the Hamiltonian parameters varies across the different dynamical regions. Despite averaging over different realizations of the disordered Ising network, the dispersion across samples (shaded in \cref{fig:results1} \textbf{a}-\textbf{b}) remains relatively small, indicating that the subsystem capacities are robust diagnostics of the underlying dynamical behaviour. Outside of the large-$W$ localised regime, our numerics are well described by an exponential fit over $f = 1,...,6$:
\begin{equation}
\label{eq:exp1}
    \mathcal{C}_f \sim \exp(\gamma f),
\end{equation}
where $\gamma$ depends on the Hamiltonian parameters. This exponential growth of Holevo capacity in subsystem size can be understood from the perspective of operator growth in this all-to-all model \cite{roberts2018operator, qi2019quantum}. In \cref{appendix:exp} we show that the leading order approximation to the Holevo quantity expanded around the \textit{Bogoliubov-Kubo-Mori} (KMB) metric \cite{petz1994geometry} not only well-approximates the full Holevo quantities presented here, but displays clearly the aforementioned connection to operator scrambling. We thus refer to $\gamma$ as an \textit{effective scrambling parameter}, since it quantifies the finite-size growth of stored input information with accessible subsystem size. For initially local injections, larger $\gamma$ indicates that recovering the stored information requires access to increasingly nonlocal degrees of freedom. We extract the scrambling parameter $\gamma$ for each sample of the Hamiltonian, and plot the results in \cref{fig:results1} \textbf{c}-\textbf{d}. 
\begin{figure}
    \centering
        \includegraphics[width=\linewidth]{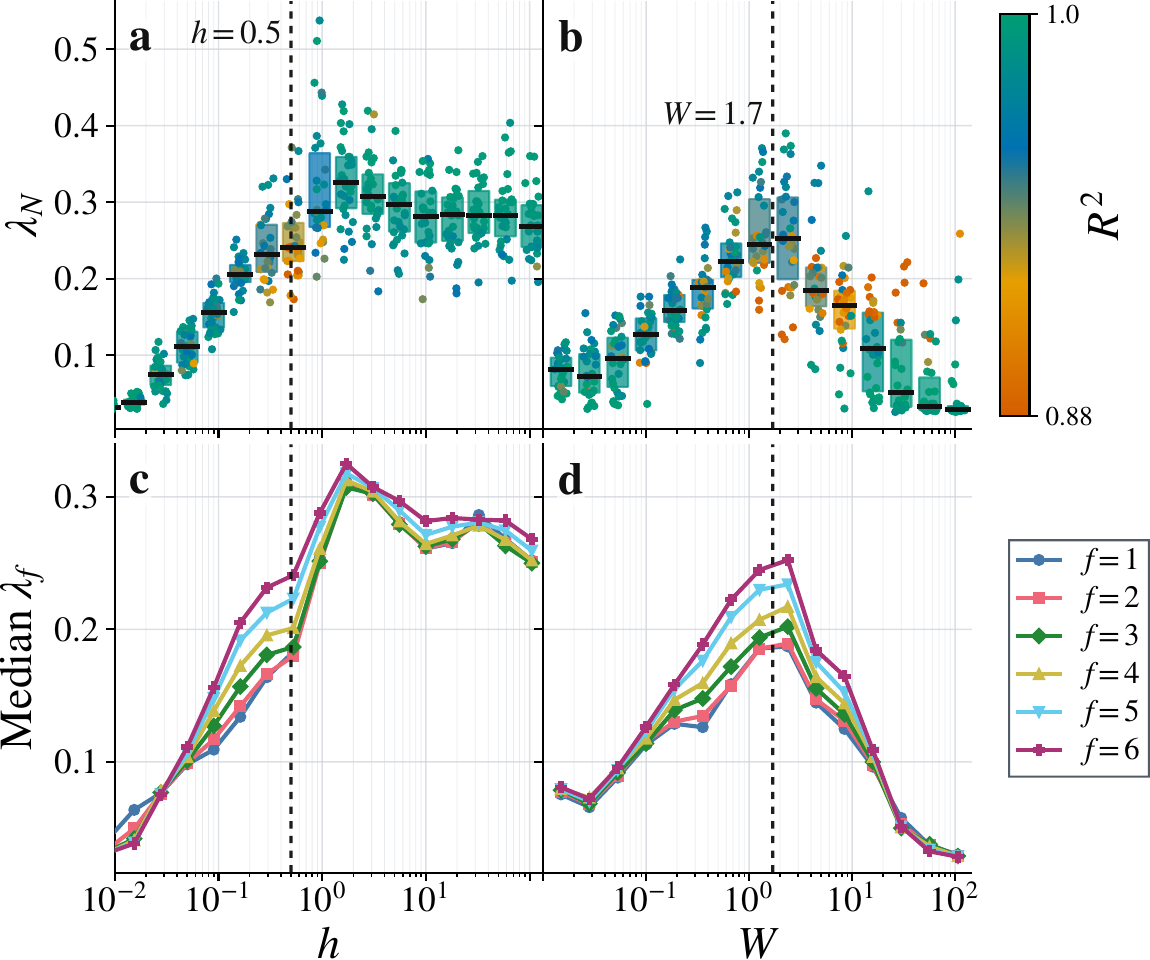}
    \caption{\textbf{a}-\textbf{b}: The extracted memory decay rate $\lambda_N$ of the full-system conditional memory $X(r,N)$ as a function of Hamiltonian parameters, with statistical presentation as in \cref{fig:results1}. \textbf{c}-\textbf{d}: Median decay rates $\lambda_f$ as a function of Hamiltonian parameters shown across different subsystem sizes. Results across $h$ values are along the cut $W=0$, while results across $W$ values are along the cut $h=0.03$.}
    \label{fig:results2}
\end{figure}
One should note that the time required for any initially local information to distribute itself across the system in this way will depend on the full system size, and so one should not expect to see this exponential behaviour in $\m C_f$ for larger system sizes with a fixed choice of $\Delta t$ \cite{yin2020bound}.

Another thing to note is that $R^2$ values should be interpreted with some caution, as the fits are performed using only six data points $f=1,\cdots,6$. For example, they remain reasonably high -- albeit noticeably lower than the average -- in the localised regime (large $W$). Still the systematic reduction of  $R^2$ in the localised regime already provides evidence for the gradual breakdown of the exponential scaling of $\m C_f$. We therefore retain the extracted parameter $\gamma$ in this regime in \cref{fig:results1} \textbf{d}, as the resulting low values for $\gamma$ still allow us to compare memory behaviour in this regime to the other parameter choices. \textcolor{black}{As a reference, consider a non-interacting reservoir with input information confined to the injection qubit. Since a fraction $f/N$ of the $f$-qubit subsystems contain this qubit, subsystem averaging gives $\m C_f=(f/N)\m C_N$. A linear fit of $\ln \m C_f$ over $f=1,\ldots,6$ yields $\gamma_{\mathrm{base}}\simeq0.343$ with $R^2\simeq0.936$. We include this baseline value of $\gamma$ in \cref{fig:results1} and highlight that moving deep into the high-$W$ localised regime brings the fitted $\gamma$ back down to this baseline value.}

It is apparent from \cref{fig:results1} \textbf{a}-\textbf{b} that the full system capacity $\m C_{f=6}$ remains relatively stable across different parameter choices, while one can notice the onset of chaos and localisation by examining the subsystem capacities. Furthermore the results for $\gamma$ in sub-panels \textbf{c}-\textbf{d} illustrate that over time, past inputs that were initially localised tend to spread nonlocally across the full system. Exponential growth in $f$ supports the interpretation that the memory $\mathcal{C}_f$ is indeed making use of many-partite storage of information across the substrate, and we see peaks in $\gamma$ corresponding to the most chaotic regions shown in \cref{fig:cuts}. As mentioned above, once the system moves into the localised regime, this exponential storage of memory in subsystem size breaks down, as transport in the system is suppressed and therefore also the distribution of the memory across the nonlocal degrees of freedom in the system \cite{serbyn2013local, huse2014phenomenology, nandkishore2015many, nico2022memory}.

\subsubsection{Fading memory}
We now move on to studying conditional Holevo quantities and the decay of information from historical inputs in the reservoir. Since $X(r,f)$ quantifies the information retained about inputs that occurred at least $r$ timesteps in the past, it is expected to decrease with $r$ as older information is progressively erased by the dissipative injection protocol. Motivated by this fading-memory behaviour, we fit the data for $1\leq r\leq 25$ with an exponential decay at the level of individual Hamiltonian realizations in order to compare the memory decay behaviour seen in \cref{fig:results0} \textbf{b} across different parameter choices:
\begin{equation}
\label{eq:exp}
    X(r,f) \sim \exp(-\lambda_f r),
\end{equation}
where the \textit{memory decay rate} $\lambda_f$ may depend on subsystem size $f$. We plot the statistics for full system decay rate $\lambda_N$ in \cref{fig:results2} \textbf{a}-\textbf{b}, while in sub-panels \textbf{c}-\textbf{d}, we plot the median decay rates $\lambda_f$ for different subsystem sizes $f$.

Firstly, we note that outside of a small window of parameters, we find that the decay rate $\lambda_f$ is approximately constant across subsystem sizes.
The differentiation of decay rates across length scales in the substrate as seen in \cref{fig:results2} \textbf{c}-\textbf{d} is most noticeable in the strongly scrambling regimes seen in \cref{fig:results1} \textbf{c}-\textbf{d}. This behaviour also seems to coincide with reduced $R^2$ values in the exponential fits used to extract $\lambda_f$. In \cref{appendix:exp} we use the KMB metric as before to assess the suitability of this exponential fit (and its subsequent breakdown in the chaotic regime) by splitting the conditional Holevo quantity into an unconditional term that inherits its decay behaviour from the action of the average map between steps, and a tripartite mutual information. As in the case of extracting the scrambling parameter $\gamma$, we find it useful to extract the \textit{effective memory decay} parameter $\lambda$ also in the regions where the exponential fit is less suitable in order to compare memory decay across Hamiltonian parameters.

As a result of the unitary invariance of the von Neumann entropy $S(U\rho U^\dagger) = S(\rho)$, one must have that both the total and conditional full system Holevo quantities are invariant under unitary conjugation. In other words, if our input injection map was unitary, we would have constant conditional Holevo quantities for past inputs. Unitary dynamics can redistribute information but cannot reduce the Holevo information associated with past inputs. While unitary evolution is responsible for scrambling and feature generation, fading memory across the full system necessarily originates from non-unitary processes. Therefore the decay of memory seen in \cref{fig:results2} is solely a result of the dissipative nature of the erase-and-write map used for injections. This map not only erases local information on the target site, but also destroys cross system correlations. On the one hand, scrambling can help to spread the information from the target site before being erased in the next step.\footnote{In fact the reason why we exclude $r=0$ in the fit \cref{eq:exp} is that in the low scrambling regions there is a large drop in conditional Holevo going from $r=0$ to $r=1$ that is not indicative of later behaviour.} On the other hand, the result of the scrambling dynamics is not that the information now lives safely on a different site of the system, but is instead spread across spatial correlations in the system that are still prone to erasure in the next step. It is therefore not obvious to what extent scrambling protects historical data from future erases, and we see in \cref{fig:results2} \textbf{a} and \textbf{b} some of the largest decay rates coincide with the regions with strongest scrambling.

\subsection{Holevo quantities \& task performance}
\label{sec:perform}
\begin{figure}
    \centering
        \includegraphics[width=0.9\linewidth]{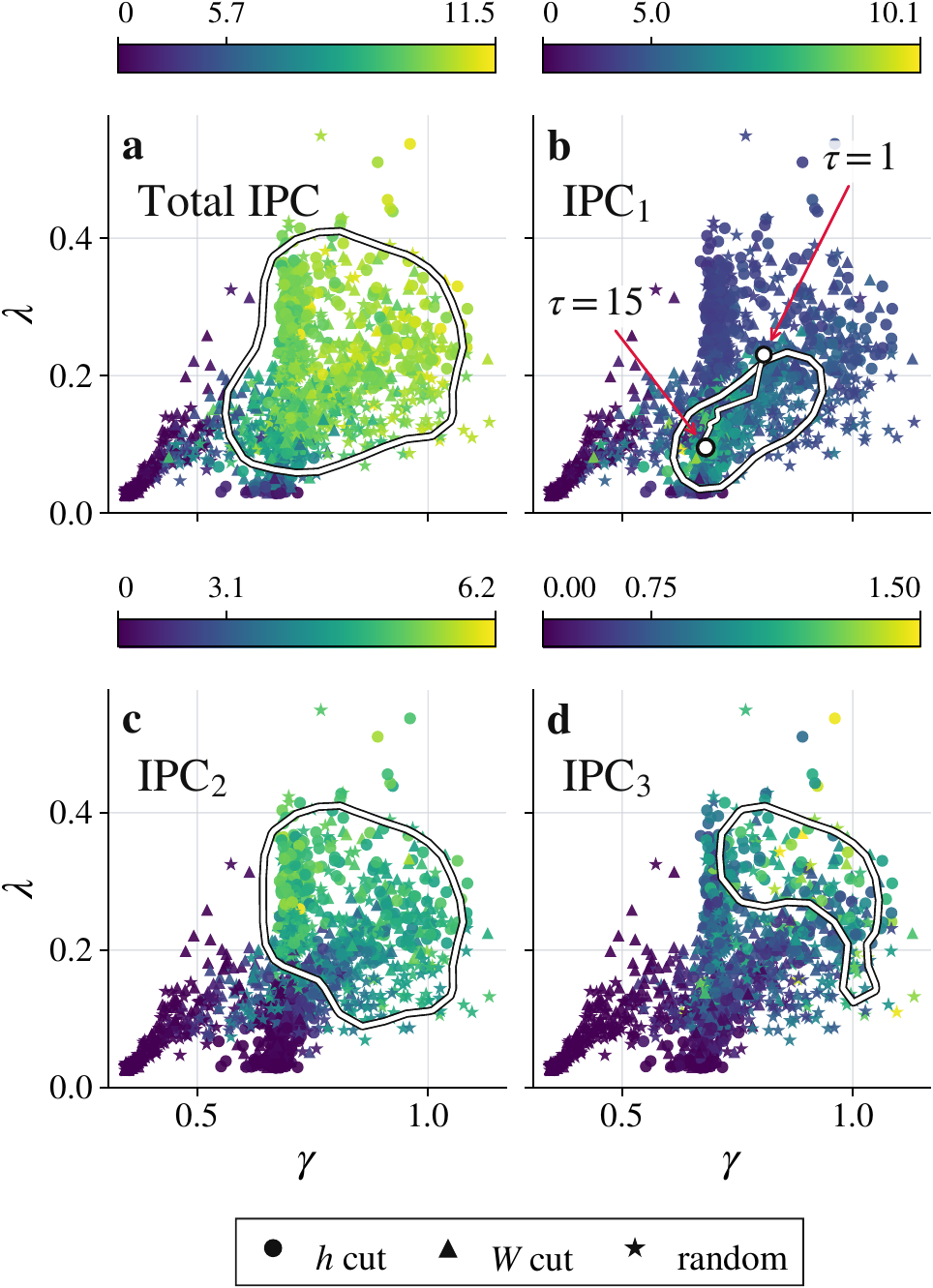}
    \caption{Total IPC capacity, along with $\mathrm{IPC}_n$ for $n=1,2,3$ against scrambling and decay parameters $\gamma$ and $\lambda$. Individual points are from a specific sampling of the Hamiltonian, and we include results from the $h$ and $W$ cuts seen in the previous figures (with $W=0$ and $h=0.03$ respectively), alongside results from random sampling of the values $10^{-2}\leq h, W\leq 10^2$. \textcolor{black}{Connected} regions of highest performance (excluding islands) are circled, and the path in panel \textbf{b} represents the region of best performance for the individual target functions with increasing delay $\tau$.}
    \label{fig:ipc}
\end{figure}
In order to relate our effective scrambling and decay diagnostics $\gamma$ and $\lambda$ to QRC performance, we calculate the information processing capacity of the protocol, or IPC \cite{dambre2012information, martinez2023information} in \cref{fig:ipc}. The IPC provides a general benchmark that characterizes the computational capabilities of a reservoir across an entire family of linear and nonlinear temporal tasks, rather than for a single application-specific benchmark. It assesses the ability of a linear readout layer, as in \cref{sec:train}, trained on measurement results, to reconstruct a family of linear and nonlinear target functions that are orthogonal with respect to the input distribution. For the evaluation of task performance, single and two-point correlators $(\langle Z_j\rangle, \langle Z_jZ_k\rangle)$ will be passed to the readout layer, with no shot noise added. 
\begin{figure*}
    \centering
        \includegraphics[width=\linewidth]{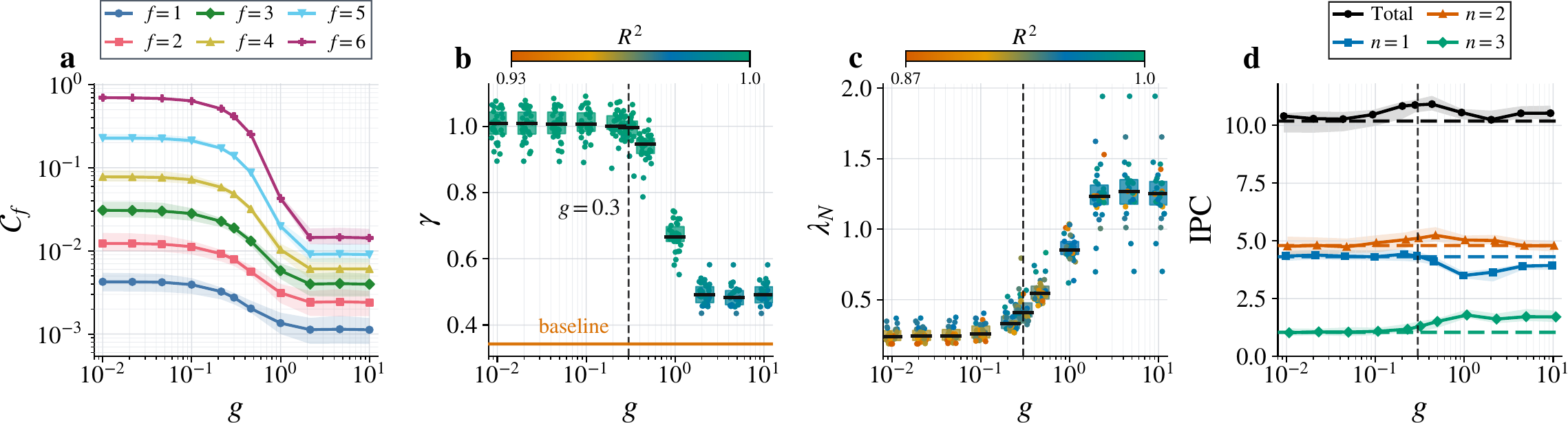}
    \caption{\textbf{a}: Memory capacity $\m C_f$ across different subsystem sizes $f$. \textbf{b}: Extracted scrambling parameter $\gamma$ using the fit \cref{eq:exp1}. \textbf{c}: Extracted memory decay parameter $\lambda$ using the fit \cref{eq:exp}. \textbf{d}: Total IPC and degree specific $\mathrm{IPC}_n$ for $n=1,2,3$ across $g$ values. The horizontal dashed lines in \textbf{d} indicate the median capacities at $g=0$, while the vertical dashed lines indicate the point $g=0.3$. All of the results in this figure are for $h=0.5$ and $W=0$. \textcolor{black}{The horizontal line labelled `baseline' in \textbf{b} shows the same non-interacting reference as in \cref{fig:results1}, with $\gamma_{\mathrm{base}}\simeq0.343$ and colour corresponding to $R^2\simeq0.936$.}}
    \label{fig:resultsg}
\end{figure*}
For the uniform distribution, the Legendre polynomials $\{\ell_k(s)\}$ form an orthonormal basis. Not only is performance tested on individual Legendre polynomials, but also on their time shifted products. For example, the target functions of degree $3$ consist not only of $\ell_3(s_t)$, $\ell_3(s_{t-1})$ etc, but also functions like $\ell_2(s_{t})\ell_1(s_{t-2})$, $\ell_1(s_t)\ell_1(s_{t-1})\ell_1(s_{t-3})$, and so on. The IPC capacity of degree $n$, or $\mathrm{IPC}_n$ is then the sum of the performances for all of the above target functions with a total degree $n$. In \cref{appendix:IPC} we discuss how performance is calculated, and add a discussion of the clustering structure seen amongst the points in \cref{fig:ipc}.

We show the total IPC, along with $\mathrm{IPC}_n$ for $n=1,2,3$ in \cref{fig:ipc}. We collect the results obtained from different $h$ and $W$ and represent the IPC as a function of the effective scrambling and memory decay rates. First of all, we find increasing total IPC capacity for increasing $\gamma$ and $\lambda$,  with a performance plateau seen in \cref{fig:ipc} \textbf{a} for larger memory decay and scrambling parameters. More interestingly, we find that the individual $\mathrm{IPC}_n$ benefit from different regions of the $\gamma$-$\lambda$ phase diagram, with best performing regions circled in each panel. In particular, we see in \cref{fig:ipc} \textbf{b} that $\mathrm{IPC}_1$ seems to benefit from less scrambling dynamics and slower decay of memory. $\mathrm{IPC}_1$ is made up of the linear target functions $y_\tau = \ell_1(s_{t-\tau}) = \sqrt{3}(2s_{t-\tau}-1)$ for increasing delay $\tau$, and we show in panel \textbf{b} that unsurprisingly, longer delays favour this weak scrambling and slow decay even further. How the encircled regions are chosen, along with the path of increasing $\tau$ in panel \textbf{b}, is detailed in \cref{appendix:IPC}.

We see that as we increase $n$ in \cref{fig:ipc} \textbf{c} and \textbf{d}, stronger scrambling and decay of memory is required for good performance. This is the so-called memory-nonlinearity trade-off that is seen across reservoir computing platforms \cite{dambre2012information, inubushi2017reservoir, vcindrak2026memory,martinez2023information}. In order for a QRC platform to be able to perform a variety of tasks, reservoir dynamics must both generate sufficiently rich nonlinear features of past inputs, and retain information about those inputs in a form accessible to the chosen measurement instrument. These requirements are often in tension, as dynamics that rapidly delocalise input information can enhance feature generation, but may also render that information less accessible to local probes. Furthermore, decay of information of past inputs may free up space in the substrate's memory not only for new inputs, but also for a wider range of nonlinearities \cite{dambre2012information}.

\subsection{Effects of finite measurement strength}
\label{sec:measurement}
Finally, we analyse the parameters $\gamma$, $\lambda$, and the IPC across different values of the measurement strength $g$, for $h=0.5$ and $W=0$. We choose this point as in \cref{fig:results1} \textbf{c}, we see that we are at a peak in the scrambling parameter $\gamma$, while at the same time in \cref{fig:results2} \textbf{c} we see that the maximal decay rate $\lambda$ has not yet been reached. Thus arises the question of the possibility of increasing reservoir performance by increasing the dissipation strength of the dynamics. The measurement strength parameter $g$ allows us exactly this control.

In \cref{fig:resultsg} we plot the subsystem capacities, scrambling and decay parameters $\gamma$ and $\lambda$, and IPC performance varying the measurement strength $g$. The first thing to note is that in \cref{fig:resultsg} \textbf{a}, the information capacities of the full- and sub-systems display a similar behaviour, dropping about an order of magnitude across the range of $g$ values investigated. We do not notice a significant corresponding drop in the IPC in panel \textbf{d}. However, given that the Holevo quantities specify how much information is at most extractable per measurement, one should expect to see the impact of this lowered capacity when shot noise is taken into consideration.

Around the value $g=1.0$, we notice a sharp transition in the behaviour of both $\gamma$ and $\lambda$ as a function of $g$. Moreover, we find that the exponential fit begins to break down (albeit not as drastically as in the large $W$ case) for $\m C_f$ in the high $g$ regime, suggesting that the dephasing effects of the large $g$ dynamics are significantly suppressing scrambling in the system, similar to the effects found in \cite{zanardi2021information}. 

For values of $g$ around $0.3$ we find that the total IPC increases compared to its $g=0$ levels, as indicated by the horizontal dashed lines in \cref{fig:resultsg} \textbf{d}. What is also interesting to note in connection to this observation, is that near $g=0.3$ one finds that the decay parameter $\lambda$ increases significantly (from a median value of $0.146$ to $0.335$), while the scrambling parameter $\gamma$ stays relatively constant (dropping from a median value of $1.00$ to $0.995$). Whereas, for example, around $g=1$ where the IPC returns to its no measurement values, $\lambda$ rises to a high value, but $\gamma$ is significantly lowered. This `sweet spot' of IPC occurs in a region where there is an increase in the decay rate in the substrate without a significant suppression of the scrambling parameter $\gamma$. Other works have already pointed out that dissipation and measurement can be used to improve performance of QRC platforms \cite{sannia2024dissipation, oriol2026,palacios2024role,mujal2023time,franceschetto2026harnessing}, and our results here suggest this to be true when the dissipation does not overly suppress scrambling.

\section{Discussion \& Outlook}
\label{sec:discussion}

In this work, we introduced a classical--quantum state that represents the inject-evolve-measure protocol commonly employed in QRC platforms, using the process tensor framework. Within the standard setting of deterministic classical inputs and single-time expectation values passed to a linear readout layer, this state collects the dependence of the information that can be passed to the readout layer on the complete input history. Mutual informations between subsets of the input history and physical subsystems of the reservoir can be expressed as Holevo quantities, providing information-theoretic diagnostics of storage, fading memory, and the local accessibility of injected information.

We applied this framework to an all-to-all transverse-field Ising reservoir with local erase-and-write injections. We found that the total information stored in the reservoir rapidly approaches a stationary value across Hamiltonian parameters, while its distribution across subsystems depends strongly on the dynamical regime. Outside the strongly localised regime, the subsystem capacities are well described, for the finite system considered here, by an exponential dependence
$\mathcal{C}_f\sim\exp(\gamma f)$. We therefore used $\gamma$ as an effective finite-size diagnostic of information scrambling: larger values indicate that access to increasingly nonlocal degrees of freedom is required to recover the stored input information. The breakdown of this behaviour in the large-disorder regime is consistent with the suppression of information transport under localised dynamics.

Conditional Holevo quantities additionally allowed us to quantify the fading memory in the substrate of historical inputs, \eqref{eq:holcond}. Their approximate exponential decay defines a memory-decay rate $\lambda$, which in the present protocol originates from the dissipative erase-and-write injection map. Scrambling can move information away from the directly overwritten site, but it also distributes that information across nonlocal correlations that remain vulnerable to subsequent erasures. The resulting relationship between scrambling and information loss is therefore not simply one of protection vs erasure: some of the largest decay rates occur in regimes where scrambling is strongest. We use the KMB approximation for the Holevo quantity in \cref{appendix:exp} in order to understand the exponential behaviour we see in both the growth of information distributed over system sizes, and the decay of information from previous inputs.

The diagnostics $(\gamma,\lambda)$ provide a useful phenomenological characterisation of QRC performance suitable to address different QRC substrates. Linear tasks favour relatively weak scrambling and slow information loss, particularly at longer delays, whereas higher-degree nonlinear tasks benefit from stronger scrambling and faster memory decay. These results give an information-theoretic perspective on the memory-nonlinearity trade-off.

The finite-measurement-strength results further illustrate this balance. Measurement-induced dephasing can increase the decay rate without substantially suppressing the scrambling diagnostic, producing an enhancement of the total IPC near $g\approx0.3$. At stronger measurement strengths, the capacity stored across both full and reduced systems decreases, the exponential subsystem-size dependence begins to break down, and scrambling is suppressed. These observations support previous findings that dissipation can improve QRC performance, while indicating that its benefit depends on increasing memory decay rates without excessively affecting the scrambling ability of the reservoir.

Several promising directions for extending the present analysis are worth noting. The scrambling diagnostic reported here is derived from a $6$-site system, with the fit spanning a modest range of subsystem sizes -- an initial estimate that motivates further investigation rather than a claim to asymptotic scaling exponents. In particular, this leaves open a natural next step: exploring the necessary conditions for the exponential spread of information at larger system sizes, for instance by allowing the inter-injection time $\Delta t$ to scale with system size, or by investigating other delocalised input injection mechanisms.

The dependence on the chosen readout observables also motivates a Heisenberg-picture formulation. The expressibility of a QRC platform is determined not only by how information is stored in the reservoir state, but by how the measured observables evolve under the reservoir dynamics. Heisenberg-picture Holevo quantities could therefore provide an observable-dependent diagnostic related to recent definitions of observable expressibility \cite{vcindrak2025krylov, vcindrak2025engineering}. Such quantities have already been used to probe scrambling in digital quantum simulators \cite{sun2025revealing}, suggesting a route towards connecting the information theoretic analysis carried out here more directly with the features available to the trained output layer.

Furthermore, in this work the linear layer is handed readout data without shot noise induced by finite sampling of the measured observables. Scrambling can reduce variations in local expectation values between input histories \cite{sannia2025exponential, xiong2025role}, and its practical effect may therefore become more restrictive when observables are estimated from a finite number of measurements. Establishing the system-size dependence of the diagnostics, comparing them with the information accessible to specified measurement schemes, and including sampling noise are important directions for future work.

A further extension of this work along the lines of the framework introduced concerns the encoding of the classical input history. The CQ state considered in this work is diagonal in the auxiliary input spaces and therefore retains only the output states associated with classical input sequences. As discussed in \cref{appendix:write}, the classical encoding map may be extended to yield a process tensor of the form
\begin{equation}
    \Upsilon^\mathrm{coh}_{1:t,F} =\int\sigma(s_{1:t},s'_{1:t})\ \mathrm{d}s_{1:t} \ \mathrm{d}s'_{1:t},
\end{equation}
where $\sigma(s_{1:t},s_{1:t})=\rho(s_{1:t})$ and the overall process tensor $\Upsilon^\mathrm{coh}_{1:t, F}$ retains positivity. Even in the case where one does not intervene in the system in an `off-diagonal' way in $s$, the above coherent extension of the CQ state may contain useful information about the extent to which different input sequences yield different output states. Such an object is used in \cite{o2026diagnosing} in order to study, for example, the dynamical entropy \cite{alicki1994defining} and spatio-temporal mutual informations in closed systems arising from a series of interventions.
\begin{figure}
    \centering
        \includegraphics[width=\linewidth]{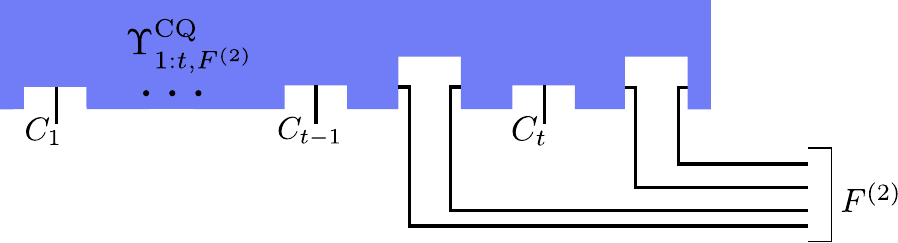}
    \caption{Process tensor for a QRC setup where general two-time instruments can be applied to the substrate.}
    \label{fig:multitime}
\end{figure}
Finally, we believe that the framework introduced here offers tools for dealing with relaxing the constraint of passing only single-time expectation values to the output layer. In the standard protocol studied here, information stored in multi-time correlations is discarded and previous measurement slots of \cref{fig:first} \textbf{b} are contracted with the average measurement channel. General multi-time instruments could instead pass temporal correlations to the output layer, and we give a schematic for a two-time setup in \cref{fig:multitime}. Bounds on information accessible through such readouts would require multi-time generalisations of the Holevo quantities, which for example have been studied recently in \cite{wu2025quantum}. Since multi-time correlations are necessary for a complete characterisation of quantum memory \cite{milz2021quantum, taranto2024characterising}, exploiting them may allow QRC architectures to access useful information-processing resources that are lacking in conventional single-time readout setups.

The CQ state framework developed here provides a first step within the commonly used single-time setting, while observable-dependent diagnostics, coherent encodings, and multi-time readouts offer routes towards characterising and exploiting more distinctly quantum forms of temporal information processing.\\
\\
\section{Acknowledgements}
We would like to thank Gian-Luca Giorgi for helpful discussions throughout the evolution of this project. N.K. would also like to thank Simon Milz and Kavan Modi for help with various aspects of the process tensor framework, and Alex Nico-Katz for helpful discussions regarding Holevo quantities.
We acknowledge funding from the Spanish State Research Agency through the COQUSY project PID2022-140506NB-C21 and -C22 and the María de Maeztu project CEX2021-001164-M, and the QuantERA QNet  PCI2024-153410 and CoQuaDis project PCI2024-153446, funded by MICIU/AEI/10.13039/501100011033 and by ERDF, EU. N.K. also acknowledges funding from the quantcom project CNS2024-15472.

\section{Data Availability}
The data and code supporting the findings of this work are available from the corresponding author upon reasonable request.

\bibliographystyle{apsrev4-2}
\bibliography{references.bib}

\appendix
\setcounter{figure}{0}
\renewcommand{\thefigure}{A\arabic{figure}}
\crefalias{section}{appendix}

\section{QRC Process Tensors}
\label{appendix:framework}
Here we will derive the CQ state \cref{eq:cq} that we use in the main text via the process tensor formalism, which is suited for describing the evolution of quantum systems in the presence of interventions.
\subsection{Choi state of the QRC process}
The Choi-Jamiołkowski isomorphism \cite{jamiolkowski1972linear, watrous2018theory} allows one to put quantum channels and higher order maps between spaces on the same footing as quantum states. Let $\m A:\m L(\m H_\textrm{i})\rightarrow \m L(\m H_\mathrm{o})$ be a quantum channel between input and output state spaces. The `Choi' state corresponding to the map $\m A$ is given by
\begin{equation}
    \mathrm A_{\mathrm{oi}}:=\mathrm{Choi}[\m A] = (\m A \otimes \mathbb I)[\Phi^+] \in \m L(\m H_\mathrm{o}\otimes \m H_\mathrm{i}),
\end{equation}
where $\Phi^+=\sum_{ij}|ii\rangle\langle jj|\in \m L(\m H_{\mathrm{i}'} \otimes \m H_{\mathrm{i}})$. The channel $\m A$ is completely positive if and only if $\mathrm A\geq 0$, and is trace preserving if and only if $\mathrm{tr}_\mathrm{o}\mathrm{A}_{\mathrm{oi}} = \mathbb
I_\mathrm{i}$.
If we want to specify that we are talking about a channel and not its Choi state, we will use script font, and if we want to specify the Choi state, we will use Roman font. We will also suppress all indices for input and output spaces when it is clear from context:
\begin{align}
    &\mathrm{A} = \mathrm{Choi}[\m A]\nonumber \\ &\mr{A_{io}} \star \rho_{\mr{i}} = \rho'_{\mr o}\nonumber \\&\mathrm{A}\star \rho= \rho' = \m A[\rho]. \nonumber
\end{align}
In the second equation above, $\star$ represents the link product \cite{chiribella2009theoretical}.
We will sometimes use the channel notation instead of the Choi in order to specify the correct ordering of compositions without having to use too many indices.

To construct the Choi state of the process tensor $\Upsilon$ corresponding to the QRC protocol (which from here on we will just refer to as the process tensor), we must first specify all of the Hilbert spaces that are involved. Of course, there are the Hilbert spaces $\m H_I$ and $\m H_F$ corresponding to the initial and final states of the system. These correspond to the global past and global future of $\Upsilon$ respectively. Since the initial state should not be relevant for the QRC due to the echo state property, we will suppress the Hilbert space $\m H_I$.

There is then an input and output Hilbert space for each intervention site where we write data to the reservoir. Given we repeat $t$ cycles, we have the `writeable' Hilbert spaces indexed by $w_{1^i},w_{1^o},\cdots w_{t^i}, w_{t^o}$. Finally, we have an input and output Hilbert space for each measurement site. For this, we have the `readable' Hilbert spaces indexed by $r_{1^i},r_{1^o},\cdots r_{(t-1)^i}, r_{(t-1)^o}$. For ease of notation, we will write $w_k = (w_{k^i}, w_{k^o})$ and $r_k = (r_{k^i}, r_{k^o})$. If $\m H_F$ is the Hilbert space of the output state, then the Choi state for this entire process up to step $t$ is the (possibly super-normalised) Choi state 
\begin{equation}
    \Upsilon_{w_{1:t}, F}^{r_{1:t-1}} \in \m L\left(\otimes_{k=1}^{t-1}[\m H_{\mathrm{w}_{k}}\otimes\m H_{\mathrm{r}_k}]\otimes \m H_{w_t} \otimes \m H_{F}\right)
\end{equation}
illustrated in \cref{fig:first} \textbf{b}.
Note that we remove the final readable slot as the final measurement can be delayed and performed on the output state on $F$.

When we do not care about the output Hilbert space (as in principle, we only pipe measurement data to the readout layer, never a quantum state), we will just write $\Upsilon_{w_{1:t}}^{r_{1:t}}$, with the final readable space added back in. Given input maps $\mr A^{(s_k)}$ for data injections, and instrument elements $\mr M^{(x_k)}$ corresponding to measurements, we have the spatiotemporal Born rule that gives the probability of getting measurement record $x_{1:t}$ given input history $s_{1:t}$:
\begin{align}
    p(x_{1:t}|&s_{1:t}) = \\ \nonumber &\Upsilon_{w_{1:t}}^{r_{1:t}} \star \mr A_{w_t}^{(s_t)} \star \mr M_{r_t}^{(x_t)} \star\cdots \star \mr A_{w_1}^{(s_1)} \star \mr M_{r_1}^{(x_1)}.
\end{align}

The most commonly used protocols do not take advantage of this full multi-time ensemble of trajectories, however, and we will now make some clarifying comments regarding which operations are effectively deterministic and which remain non-deterministic in the context of common QRC protocols.

\subsection{Deterministic vs. non-deterministic operations \& the QRC process tensor}
Instruments provide a formalism for implementing non-deterministic operations: an instrument $\m J$ is an ensemble of CP maps $\{\m M^{(x)}\}$ that act non-deterministically on the state. In particular, an application of the instrument $\m J$ on the state $\rho$ yields one of the states $\rho^{(x)} = \m M^{(x)}(\rho)$, with probabilities $p_x = \mathrm{tr}\rho^{(x)}$. This can be written as a deterministic map by introducing a classical auxiliary space $\m H_\mathrm{cl}$:
\begin{align}
    \mathcal{J}&:\m L(\m H_\mathrm{in})\rightarrow \m L(\m H_\mathrm{out}\otimes \m H_\mathrm{cl}) \\ &:\rho\rightarrow \sum_x \m M^{(x)}[\rho]\otimes |x\rangle\langle x|,
\end{align}
where the non-determinism is implemented by projecting the above state onto a particular choice of classical outcome $x$.

Instruments are thus commonly used for representing the effects and outcomes of (sharp and unsharp) measurements. A single shot outcome of measuring $\rho$ will yield result $x$ with probability $p_x$, and result in the (unnormalised) post-measurement state $\rho^{(x)}$. Over many shot outcomes of the same repeated non-deterministic process, an average state $\sum_x \m M^{(x)}[\rho]$ emerges, where $\tilde{\m M} = \sum_x \m M^{(x)}$ is a CPTP map called the average measurement channel.\\
\\

\textbf{Comment 1:} As mentioned in the main text, the typical QRC setup in the literature has a linear output layer trained only on single-time expectation values. In other words, experimental statistics making up multi-time correlations are not gathered. We therefore lose any information contained in trajectories induced by previous measurements, the result of which is that regarding information extracted from the QRC at time $t$, measurement slots for $k<t$ are always effectively contracted with the (deterministic) average measurement channel $\tilde{\m M}$.\\
\\

\textbf{Comment 2.} If one considers the input time series as being generated by a random process then it might make sense to model the data writing via non-deterministic input maps. However, there is a contrast between the way in which measurements are non-deterministic, and the way in which the input time series is non-deterministic. Consider a typical QRC experiment picture: a time series to be processed is sampled and used as input for the QRC. Expectation values are needed, and so the whole QRC evolution with the same time series input is repeated many times. Each shot of the experiment yields different measurement outcomes, but the same time series is used as input. In this sense, the input time series is deterministic with respect to these different trials of the experiment. This is not to say that different time series cannot be used in different batches of experiments. In fact if one wants to test the separation or fading memory properties, then different samplings of the time series are indeed needed. However, within each experiment batch the input maps are deterministic. Therefore we will be using CPTP maps to model the data injection maps as opposed to instruments. 

With all this in mind, we will now introduce the QRC process tensor for the reservoir at step $t$ for deterministic injections with single-time expectation values passed to the readout layer:
\begin{equation}
    \Upsilon _{w_{1:t}, F} := \Upsilon_{w_{1:t}, F}^{r_{1:t-1}}\star \tilde{\mr M}^{\otimes (t-1)}_{r_{1:t-1}},
\end{equation}
where at each measurement slot we assume the same instrument $\m J=\{\m M^{(x)}\}$ with average measurement Choi state $\tilde{\mr M} = \mathrm{Choi}(\sum_x \m M^{(x)})$. The tensor $\Upsilon_{w_{1:t}, F}$ contains all of the information that could possibly be sent to the output layer at time $t$.\footnote{Even though we write $\Upsilon_{w_{1:t}, F}$ to mean the process tensor for inputs $1$ to $t$, we may choose when we care to start `counting' the inputs, usually assuming that we are already beyond the washout time so that the QRC dynamics is independent of any effects particular to a choice of initial state.} 

From here onwards, we will avoid writing the average measurement map $\tilde{\m M}$ explicitly, and just absorb it into the dynamical map $\m E$.
\begin{figure}
    \centering
        \includegraphics[width=\linewidth]{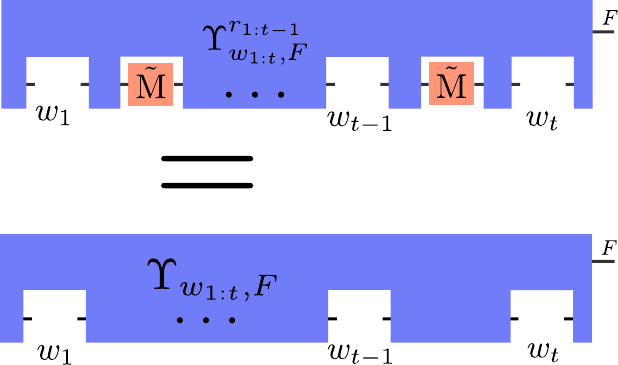}
    \caption{The `write-only' process tensor one gets by contracting all previous measurement slots in $\Upsilon_{w_{1:t}, F}^{r_{1:t-1}}$ with the average measurement map $\tilde{\mr M}$.}
    \label{fig:distev}
\end{figure}
\subsection{Writeable spaces for the QRC process tensor}
\label[appendix]{appendix:write}
Next, we will investigate the writeable slots in $\Upsilon _{w_{1:t}, F}$. For the moment, let us consider the single-slot tensor $\Upsilon_{w_1, F}$ with global future (Hilbert space for output state). The single input $s$ is written to the reservoir by inserting the Choi state of some CPTP map $\m {A}_1^{(s)}$ into the single slot of the tensor:
\begin{equation}
    \Upsilon_{w_1, F} \star \mathrm{A}_1^{(s)} = \rho(s),
\end{equation}
where $\rho(s) = \m E\circ  \m A^{(s)} [\rho]$ for some choice of initial state $\rho$.

We will need to define a basis of CP maps\footnote{Note that not even a CPTP basis will work here: if we have a continuous family of unitary interventions no orthogonal CPTP basis will exist since unitaries are extremal points of the convex set of CPTP maps.} that we can expand our input maps over. Following \cite{milz2017introduction}, we define such a basis $\{\mr B_j\}$ in the Choi representation, along with a dual frame $\{\mr D_j\}$. If we fix the pattern of the inject-evolve cycle as before, then we may write the tensor as:

\begin{equation}
    \Upsilon_{w_1, F} = \sum_j (\m E\circ \m B_j)[\rho]\otimes \mr D^*_j.
\end{equation}
Then by expanding the input maps as:
\begin{equation}
    \mr A^{(s)} = \sum_k \alpha_k^{(s)} \mr B_k,
\end{equation}
and inserting $\mr A^{(s)}$ into the single intervention slot, we get:
\begin{align}
    \Upsilon_{w_1, F} \star \mr A^{(s)} &= \sum_{jk} \alpha^{(s)}_k (\m E\circ \m B_j)[\rho_0]\otimes \mathrm{Tr}(\mr D^*_j \mr B_k^T) \nonumber \\ 
    &= \sum_{jk} \alpha^{(s)}_k\delta_{jk} (\m E\circ \m B_j)[\rho_0] \nonumber \\
    &= \left(\m E \circ \sum_k \alpha^{(s)}_k \m B_k\right)[\rho_0] \nonumber \\ 
    &= (\m E \circ \m A^{(s)})[\rho_0] \nonumber \\ &=\rho(s),
\end{align}
as required. The generalisation to the multi-slot write-only tensor is straightforward and we write the explicatory equalities here for completion:
\begin{align}
    &\Upsilon_{w_{1:t}, F}\star \mr A^{(s_t)} \star\cdots\star \mr A^{(s_1)} \nonumber \\
    &= \big(\m E \circ \m A^{(s_t)}\circ  \cdots \circ \m E\circ \m A^{(s_1)}\big)[\rho_0]\nonumber \\ & = \rho(s_1,\cdots , s_t) = \rho(s_{1:t}).
\end{align}

\subsubsection{The classical encoding map}
\label{sec:classical}
\begin{figure}
    \centering
        \includegraphics[width=\linewidth]{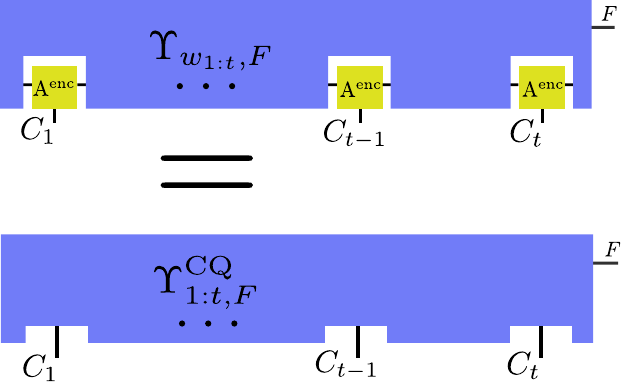}
    \caption{Classically controlled process tensor one gets by contracting all input slots of the tensor $\Upsilon_{w_{1:t}, F}$ with the classical encoding map $\mr{A}^{\mr{enc}}$.}
    \label{fig:enc}
\end{figure}
We may want not only to restrict the CP basis for the input spaces in order to better understand a given QRC setup choice, but we also want to consider the Choi state we get when we contract the process tensor with an already chosen method of writing information into the system. The result of this in the case of classical inputs $s_{1:t}$ is that our tensor $\Upsilon_{1:t, F}$ reduces to the classical-quantum (CQ) state:
\begin{equation}
    \Upsilon_{1:t,F}^\mathrm{CQ} = \sum_{s_{1:t}} \rho(s_{1:t}) \otimes |s_{1:t}\rangle\langle s_{1:t}|.
\end{equation}
To see this, we introduce the classical auxiliary space $\m H_C$ for each intervention slot, and construct the classically controlled (unnormalised) encoding map:
\begin{equation}
\label{eq:enc}
   \mr A^\mathrm{enc}_{\mr w, C} = \sum_s \mr A^{(s)}_{\mr w}\otimes |s\rangle\langle s|_C,
\end{equation}
which can be understood as follows: Given a classical space $\m H_C$ with dimension such that all possible classical inputs we could apply form an orthonormal basis $\{|s\rangle\langle s|_C\}$, we can consider $\mr A^\mathrm{enc}_{\mr w, C}$ to be controlled on this space. Given a classical input $|s\rangle\langle s|_C$, $\mr A^\mathrm{enc}_{\mr w, C} $ returns the map $\mr A^{(s)}_{\mr w}$. Using our CP basis from earlier $\{\mr B_j\}$, we can write (with suppression of indices)
\begin{equation}
   \mr A^\mathrm{enc} = \sum_{s }\mr A^{(s)} \otimes |s\rangle \langle s| = \sum_{sk}\alpha_k^{(s)}\mr B_k\otimes |s\rangle \langle s|.
\end{equation}
Contracting this with the single slot write-only tensor, we get:
\begin{align}
    \Upsilon_{w_1, F} \star \mr A^\mathrm{enc}_{w_1} &= \sum_{sjk} \alpha^{(s)}_k \mathrm{tr}(\mr D_j^*  \mr B_k^T)\m (\m E\circ \m B_j)[\rho_0]\otimes |s\rangle\langle s|\nonumber \\ &=\sum_{s} \m E\circ \m A^{(s)}[\rho_0] \otimes |s\rangle\langle s| \nonumber \\ &=\sum_s \rho(s)\otimes |s\rangle\langle s|,
\end{align}
which again, can be easily generalised to the multi-injection case:
\begin{equation}
    \Upsilon_{1:t,F}^\mathrm{CQ} = \Upsilon_{w_{1:t}, F} \star (\mr{A}^\mathrm{enc})^{\otimes t} = \sum_{s_{1:t}} \rho(s_{1:t}) \otimes |s_{1:t}\rangle\langle s_{1:t}|.
\end{equation}
This CQ state has a straightforward interpretation: it has a classical register in $|s_{1:t}\rangle\langle s_{1:t}|$ that records the past inputs to the reservoir, and a quantum register $\rho(s_{1:t})$ that contains the resulting states of those past inputs. Contracting with a choice of past inputs $s'_1,\cdots, s_t'$ yields the state
\begin{equation}
    \Upsilon_{1:t,F}^\mathrm{CQ} \star |s'_{1:t}\rangle \langle s'_{1:t}| = \rho(s'_{1:t}).
\end{equation}

\subsubsection{Off diagonal coherences in $\mr{A}^{\mr{enc}}$}
\label[appendix]{appendix:coh}
The mutual information between intervention spaces and final output space of a process tensor is approached in other works that study, for example, the scrambling behaviour of closed systems \cite{dowling2024operational, o2026diagnosing}. Although not focused on quantum reservoir computing, this literature is quite relevant to examining QRC protocols as the authors use process tensors to study how the information injected into a closed system via unitary (or rank-1 projection) perturbations gets scrambled by the unitary propagator of the system, leading to novel diagnostics for chaotic behaviour. 

Consider a QRC platform that encodes classical data $s$ in the reservoir by application of a unitary $U^{(s)}$:
\begin{equation}
    \mr{A}^{\mr{enc}} \star \rho_0 = \sum_s U^{(s)}\rho_0 U^{(s)\dagger}\otimes |s\rangle\langle s|,
\end{equation}
where $U^{(s)}$ may act non-trivially only on a subsystem. We can reproduce the pure process tensor studied in \cite{o2026diagnosing} by enforcing a pure initial state, requiring $\m E$ to be unitary, and adding off-diagonal terms in the classical basis of $\mr{A}^{\mr{enc}}$. This `coherent' extension of the encoding map can be written as:
\begin{equation}
   \mr{ A}^{\mr{coh}} \star \rho_0 = \sum_{ss'} U^{(s)}\rho_0 U^{(s')\dagger}\otimes |s\rangle\langle s'|,
\end{equation}
which maintains the complete-positivity of diagonal-in-$s$ version of $\mr A_\mathrm{enc}$. The corresponding coherent-in-$s$ process tensor is then given by
\begin{align}
    \Upsilon_{1:t,F}^\mathrm{coh} &= \sum_{s_{1:t},s'_{1:t}} \m E(U^{(s_t)}\m E(\cdots \label{eq:wmany}\\&\m E(U^{(s_1)}[\rho_0]U^{(s_1')\dagger})\cdots)U^{(s_t')\dagger}) \otimes |s_{1:t}\rangle\langle s'_{1:t}|\nonumber.
\end{align}
Since we are assuming unitary $\m E$ and pure initial state, we can then simplify this expression by defining the pure state $|\psi(s_{1:t})\rangle$ to be the result of an intervention history of $s_{1:t}$:
\begin{align}
    \Upsilon_{1:t,F}^\mathrm{coh} = \sum_{s_{1:t},s'_{1:t}}|\psi(s_{1:t})\rangle\langle \psi(s'_{1:t})| \otimes |s_{1:t}\rangle\langle s'_{1:t}|.
\end{align}
For phase encoding on a single qubit via $U^{(s)} = R_z(s)$ with $s\in\{0, \pi\}$, this is exactly the object studied in \cite{o2026diagnosing} for unitary interventions. Tracing over the global future (output space) of the tensor gives us a reduced (super-normalised) Choi state that encodes coherences between different input histories via fidelities between the output states of the process:
\begin{align}
    \mathrm{tr}_F(\Upsilon_{1:t,F}^\mathrm{coh} ) \nonumber= \sum_{s_{1:t},s'_{1:t}}\langle \psi(s'_{1:t})| \psi(s_{1:t})\rangle |s_{1:t}\rangle\langle s'_{1:t}|.
\end{align}
The authors of \cite{o2026diagnosing} then use (the normalised version of) this reduced state to define the \textit{quantum dynamical entropy} of the intervention process, among other diagnostics, which are then used to investigate the scrambling behaviour of chosen closed system propagators. More specifically, they use the many-intervention behaviour of this object:
\begin{equation}
    S^\mathrm{dy}(t):=\frac{1}{t}S\left(\frac{1}{d^t}\mathrm{tr}_F\Upsilon_{1:t,F}^\mathrm{coh}\right).
\end{equation}
This construction of the coherences-in-$s$ version of the process tensor $\Upsilon_{1:t,F}^\mathrm{coh}$ can be carried out\footnote{Although not without having to choose the way in which these coherences are generated.}, also in the cases of non unitary $\m E$, and non rank-1 interventions $\mr {A}^{(s)}$, and therefore applied to our general QRC framework. This also holds regarding the more general \textit{spatio-temporal mutual informations} studied in the above mentioned text, the corresponding diagonal-in-$s$ versions being Holevo quantities restricted to both intervention and physical subsystems. 

Although for QRC experiments with classical inputs, one never injects data in this `off-diagonal' way, it could be useful to investigate these coherent extensions of $\mr A^{\mr{enc}}$ and the corresponding entropic quantities for their abilities to study information scrambling in the reservoir substrate. For this work however, we focus on the diagonal-in-$s$ encoding maps, and so stick with the corresponding Holevo quantities.

\subsubsection{Continuously-valued inputs}
In the above exposition, we assumed discretely-valued inputs for $s_{1:t}$. This was for ease of notation, and in fact the majority of work on studying QRC platforms assumes instead continuously-valued inputs, as do we in the main text of this work. However, the inputs for the setups in this work are sampled from the compact interval $\Omega = [0,1]$, and all of the results from this appendix are easily generalisable to such a compact distribution using the uniform independent measures
\begin{equation}
    \mathrm{d}s_{1:t} = \prod_{k=1}^t\ \mathrm{d}s_{k}.
\end{equation}

\section{Discussion and motivation of exponential fits}
\label{appendix:exp}
In order to comment on the fits used in the main text, we first introduce a useful approximation of the Holevo quantity.
\subsubsection{KMB approximation}
Similar to \cite{ding2026thermodynamics}, we use the Kubo-Mori-Bogoliubov (KMB) approximation for the Holevo quantity. Given an ensemble of states $\{p(x), \rho(x)\}$, with $p$ classical probability distribution, so long as the support of the individual states $\rho(x)$ is contained within the support of the average state $\overline \rho = \int \rho(x) \ p(x) \mr dx$, the Holevo quantity can be expanded around a quadratic form in the state fluctuations:
\begin{align}
    \chi(\{p(x), \rho(x)\}) &= S\left(\overline \rho\right) - \int S(\rho(x)) \ p(x) \mr d x \\ &= \int D(\rho(x) || \overline \rho) \ p(x)\mr dx \\ &\approx \frac{1}{2} \int g^{\mathrm{KMB}}_{\overline \rho}(\delta \rho(x), \delta \rho(x) )\ p(x)\mr dx, \label{eq:KMB}
\end{align}
where $g^{\mathrm{KMB}}_{\overline \rho}$ is the KMB metric \cite{petz1994geometry} on the tangent space at $\overline\rho$, $\delta \rho(x) = \rho(x) - \overline \rho$, and the final line is an approximation up to terms that are third order in the state fluctuations. Thus the above approximation holds well for situations when typical fluctuations around the average state remain small. In \cref{fig:trial} we plot this KMB approximation against our full Holevo results for individual samples of the Hamiltonian in \cref{eq:hamil}, and find tight agreement.

\subsubsection{Exponential growth of $\m C_f$}
\begin{figure}
    \centering
    \includegraphics[width=\linewidth]{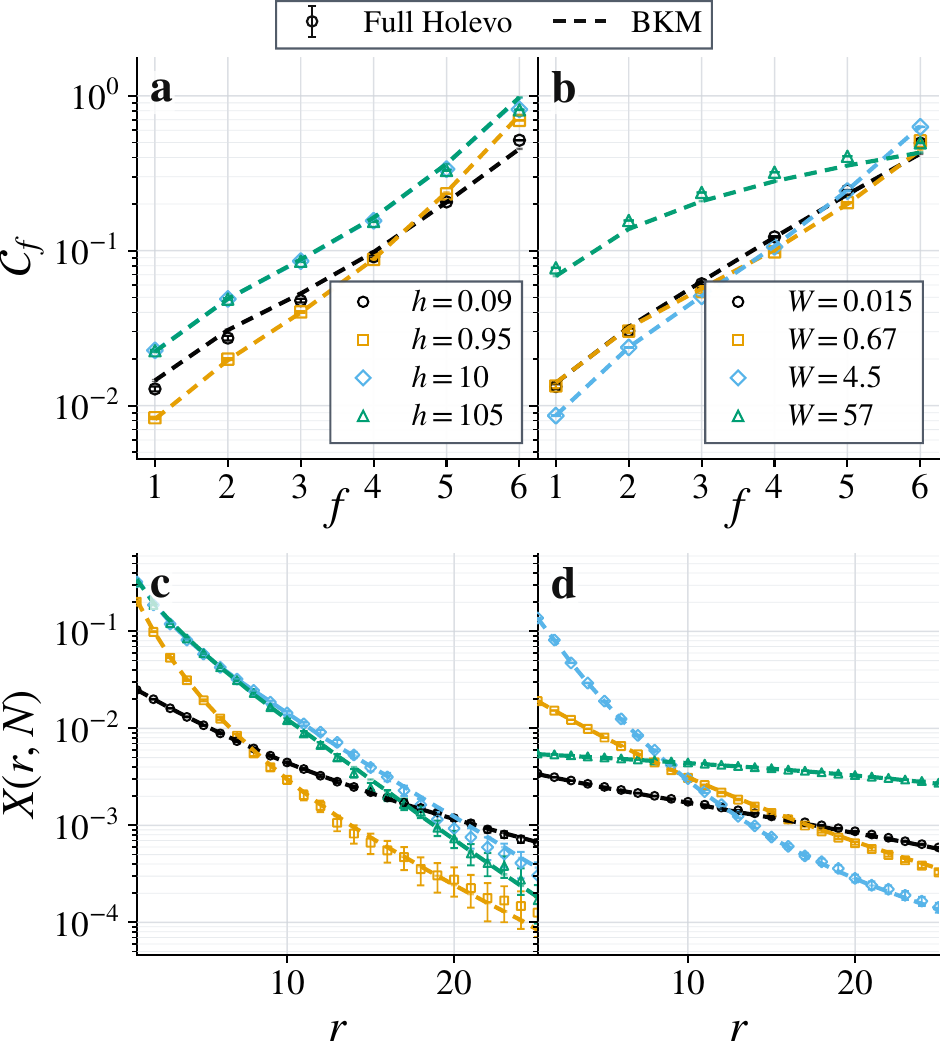}
    \caption{\textbf{a}-\textbf{b}: Subsystem capacities $\m C_f$ as a function of subsystem size. \textbf{c}-\textbf{d}: Conditional Holevo quantities as a function of conditioned window width $r$. Results across $h$ values (\textbf{a}, \textbf{c}) are along the cut $W=0$, while results across $W$ values (\textbf{b, d}) are along the cut $h=0.03$. The results are shown for a particular sampling of the Hamiltonian \cref{eq:hamil}, with error bars to indicate SEM in calculating the Holevo quantity via Monte-Carlo sampling of input sequences. All results are taken for Holevo values at $t=50$. Markers with error bars indicate full Holevo results, while dashed lines indicate the KMB approximation of the Holevo quantities.}
    \label{fig:trial}
\end{figure}

In \cref{fig:trial} \textbf{a}-\textbf{b} we plot $\m C_f$ -- the long time behaviour of the Holevo quantity over all past inputs, averaged over subsystems consisting of $f$ physical units. Consider bipartitions of the physical system $F=F_1\otimes F_2$, so we can write
\begin{equation}
    \m C_f = \frac{1}{{N\choose f}}\sum_{|F_1| = f}\m C_{F_1}.
\end{equation}
Using \cref{eq:KMB},
\begin{equation}
    \m C_{F_1} \approx \frac{1}{2}\int g^{\mathrm{KMB}}_{\overline \rho_{F_1}}(\delta \rho_{F_1}(s_{1:t}), \delta \rho_{F_1}(s_{1:t})) \ \mathrm ds_{1:t},
\end{equation}
where $\rho_{F_1} = \mathrm{tr}_{F_2}\rho$. Then defining the set of Pauli strings on $f$ sites as $\mathcal{P}_f = \{\mathbb I, X,Y,Z\}^{\otimes f}$, we can expand the above expression in the Pauli basis as
\begin{align}
&\frac{1}{2d_f^2}\sum_{P,Q\in \mathcal{P}_f\backslash\mathbb{I}} g^{\mathrm{KMB}}_{\overline \rho_{F_1}}(P, Q) \int \delta c^{F_1}_P(s_{1:t}) \delta c^{F_1}_Q(s_{1:t}) \ \mathrm ds_{1:t}\nonumber \\ = &\frac{1}{2d_f^2}\sum_{P,Q\in \mathcal{P}_f\backslash\mathbb{I}} g^{\mathrm{KMB}}_{\overline \rho_{F_1}}(P, Q) \  \mathrm{Cov}[c^{F_1}_P(s_{1:t}),c^{F_1}_Q(s_{1:t})]
\label{eq:expKMB}
\end{align}
where $c^{F_1}_P(s_{1:t}) = \mathrm{tr}(\rho_{F_1}(s_{1:t}) P)$. 
Due to the sum over the Pauli operator basis, the number of terms in the above expression grows exponentially in $f$, although this does not automatically imply that $\mathcal{C}_f$ will grow exponentially and the dynamics of the reservoir plays a key role in determining how many operators end up meaningfully contributing to $\mathcal C_f$. In particular, the question of how many such terms in \cref{eq:expKMB} end up being relevant is closely related to the mechanism of operator spreading in quantum many-body systems \cite{nahum2018operator, qi2019quantum}. An initial operator propagated by the dynamics of the system can become spread across independent directions in the operator space, even in integrable systems \cite{gopalakrishnan2018hydrodynamics}, while in localised systems this operator spreading is suppressed \cite{chen2017out, swingle2017slow}. As can be seen in \cref{eq:expKMB}, this operator spreading (or suppression thereof) should -- via influencing the variances of coefficients in a smaller or larger subset of Pauli strings -- to a significant extent determine the information propagation within the system, a conclusion that is supported by recent works \cite{shang2025operator}. Since the model we are analysing in this work has an all-to-all connectivity, spatial profiles such as lightcones are replaced by operator size distributions \cite{roberts2018operator, qi2019quantum}, and we suggest that this operator growth can be understood as the mechanism behind why we observe exponential growth of $\m C_f$ in $f$, along with why we see the breakdown of this behaviour in the localised regime.

\subsubsection{Exponential decay of conditional Holevo quantities}
Other works note the exponential decay of memory in QRC platforms \cite{sannia2026non}, and in \cref{fig:trial} \textbf{c}-\textbf{d}, we plot the decay of the conditional Holevo quantity $X(r,f)$ in $r$. As discussed in the main text, we see the biggest deviation from exponential decay in the regions of strongest scrambling, which in panel \textbf{c} can be seen for $h=0.95$, and in panel $\mathbf{d}$ for $W=4.5$.

\begin{figure}
    \centering
        \includegraphics[width=\linewidth]{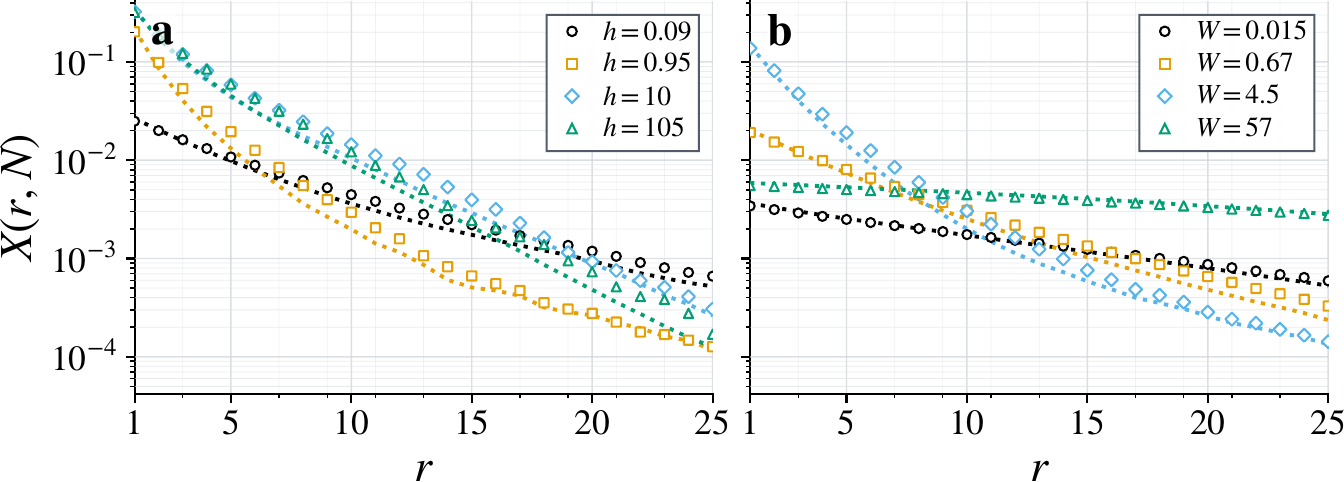}
    \caption{\textcolor{black}{Conditional Holevo quantities $X(r,N)$ (markers) and unconditional mutual information $I(C_H;F)$ (dashed lines) for a single Hamiltonian realisation. Panel \textbf{a} shows results across $h$ values for $W=0$, while panel \textbf{b} shows results across $W$ values for $h=0.03$.
}}
    \label{fig:unc}
\end{figure}

We first note that the QMI that gives the conditional Holevo used in the main text can be split into two contributions:
\begin{equation}
\label{eq:unc}
    I(C_H; F | C_R) = I(C_H ; F) - I_3(C_H ; C_R ; F),
\end{equation}
an unconditional contribution $I(C_H ; F)$ that ignores information from inputs $R$, and the tripartite mutual information $I_3(C_H ; C_R ; F)$ \cite{mcgill1954multivariate} that contains information spread across all three spaces. The first term in the decomposition can be understood as an unconditional Holevo quantity over the historical inputs from $H$:
\begin{equation}
    I(C_H ; F)  = \chi_t^\mathrm{unc}(r) = \int_{\Omega^{t-r}} D(\overline \rho_t^R(s_H) || \overline \rho_t) \ \mr ds_H,
\end{equation}
where $\overline \rho_t^R(s_H)$ is the state at step $t$ averaged over recent inputs $s_R$ for fixed choice $s_H$ of historical inputs, while $\overline \rho_t$ is the state at step $t$ averaged over all previous inputs. We can relate the decay of $\chi_t^\mathrm{unc}(r)$ to the slowest decaying mode of the average map between steps using the KMB approximation from earlier. Given that the reservoir evolves with the dynamical map $\m E$ in between steps, and the average injection map is given by $\overline{\m A}$, we have the full average map $\Lambda = \m E \circ \overline{\m A}$, and so:
\begin{equation}
    \overline \rho_t^R(s_H) = \Lambda^r[\rho(s_H)]. 
\end{equation}
If we let $\sigma$ be the steady state and $R_k$ be the right decaying eigenmodes of $\Lambda$ with eigenvalues $\mu_k$, we can write:
\begin{align}
    \rho(s_H) &= \sigma + \sum_k a_k(s_H) R_k \\ 
    \overline \rho_t^R(s_H) &= \sigma + \sum_k \mu_k^r a_k(s_H) R_k.
\end{align}
Then using this expansion around the steady state $\sigma$, we can use the KMB approximation to write
\begin{equation}
    \chi_t^\mathrm{unc}(r) \approx \frac{1}{2}\sum_{jk}(\mu_j\mu_k^*)^rg^\mathrm{KMB}_{\sigma}(R_j, R_k)\mathrm{Cov}[a_j(s_H), a_k(s_H)].
\end{equation}
We see immediately that $\chi_t^\mathrm{unc}(r)$ will exhibit exponential decay if there is a dominating decay mode in the above decomposition of $\rho(s_H)$. On the other hand, a breakdown of the exponential behaviour in the decay of the conditional Holevo quantity $\chi_t(r)$ can thus either arise from a competition between decaying modes in the aforementioned expansion, or a competition between $\chi_t^\mathrm{unc}(r)$ and the tripartite mutual information $I_3(C_H ; C_R ; F)$. We plot both the conditional and unconditional Holevo quantities in \cref{fig:unc} \textbf{a}-\textbf{b}, and see that this breakdown in exponential decay in the high scrambling regime also occurs for the unconditional quantity. \textcolor{black}{The similar deviations from exponential decay in the conditional and unconditional quantities are consistent with contributions from multiple decay modes of the average map $\Lambda$ in the strongly scrambling regime.}

\section{Extra numerical details}
\label{appendix:numerical}
\subsection{Simulation parameters}
\label[appendix]{appendix:params}
\begin{figure}
    \centering
        \includegraphics[width=\linewidth]{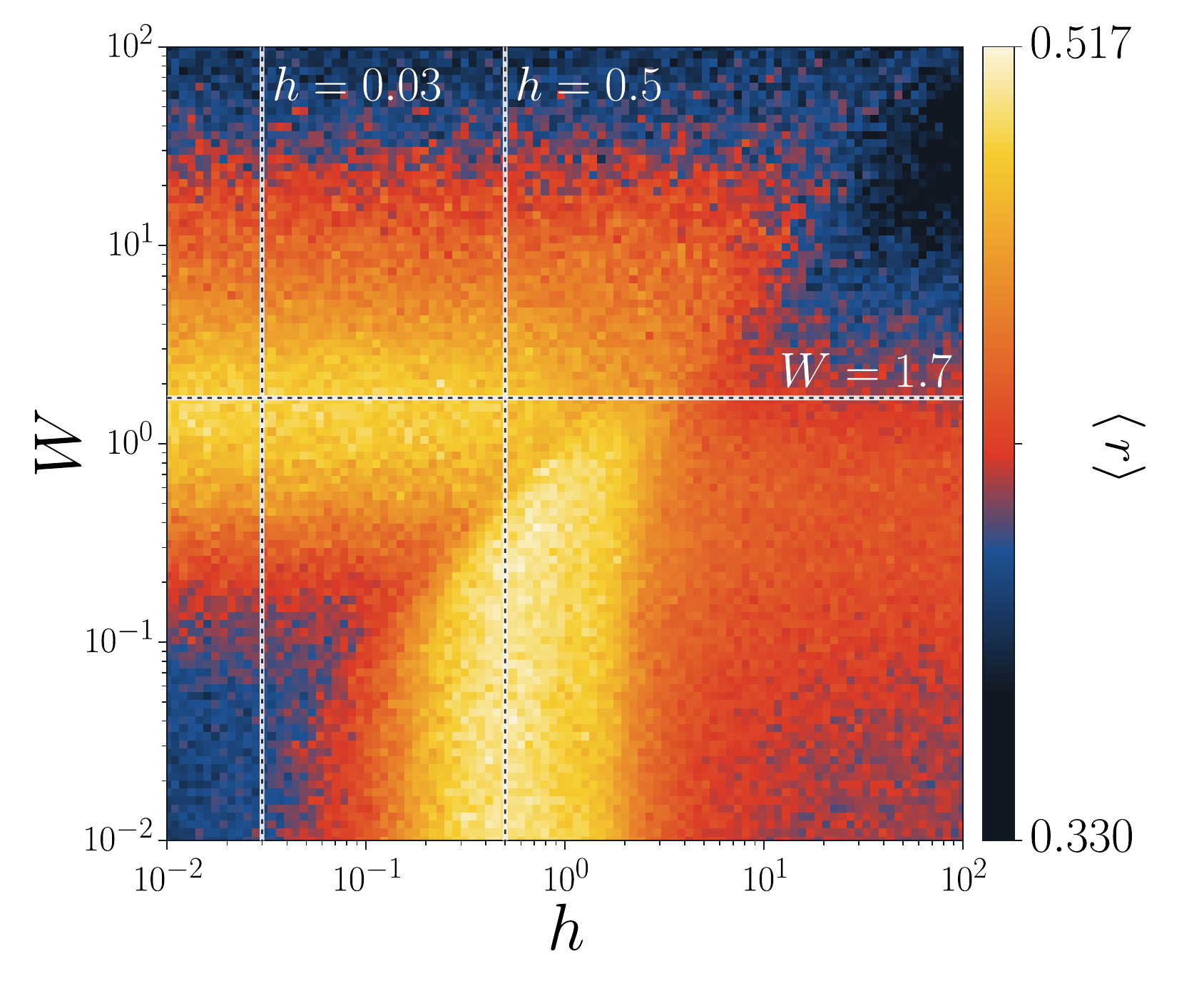}
    \caption{Phenomenological phase diagram of the Hamiltonian \cref{eq:hamil} for $N=6$ sites, using the symmetry-resolved mean level spacings $\langle r\rangle$. Parameter values of interest are marked with dashed lines.}
    \label{fig:phase}
\end{figure}
Here we give a summary of the details of the setup used for the numerics in this work.
\begin{itemize}
    \item We take the time between injections to be $\Delta t = 10$.
    \item Before calculating anything, we initialise the reservoir in the all zero state, and then evolve  the reservoir with a random string of inputs for a washout time of $1000$ steps.
    \item We then analyse Holevo quantities over a further window of $50$ steps.
    \item Conditional Holevo quantities of historical inputs are calculated for recent input window widths from $r=1$ to $r=25$.
    \item All results (aside from those that come from the random selection of $h$ and $W$ used in \cref{fig:ipc}) are taken over $30$ realisations of the random Hamiltonian in \cref{eq:hamil}. We plot a phenomenological phase diagram using mean level spacing statistics for the Hamiltonian in \cref{fig:phase}
\end{itemize}

Now we will describe how the Holevo quantities corresponding to continuously-valued inputs are calculated. Recall that the full Holevo quantity over the continuous ensemble $\{p(s_{1:t}), \rho(s_{1:t})\}$ is given by
\begin{equation}
    S\left(\int \rho(s_{1:t}) \ \mr{d}s_{1:t}\right) - \int S(\rho(s_{1:t})) \ \mr d s_{1:t}.
\end{equation}
The second term is calculated by Monte-Carlo sampling the trajectories, and in all of the results in this work we take $10,000$ samples. The first term can be calculated by averaging the input map, which in the case of our erase-and-write map can be done exactly. A similar procedure is used to also calculate the conditional Holevo quantities, and in \cref{fig:trial} we show the full Holevo and conditional Holevo quantities for a specific sample of the Hamiltonian \cref{eq:hamil}, with error bars from the sampling procedure detailed above.

\subsection{Further IPC details}
\label[appendix]{appendix:IPC}
\begin{figure}
    \centering
        \includegraphics[width=\linewidth]{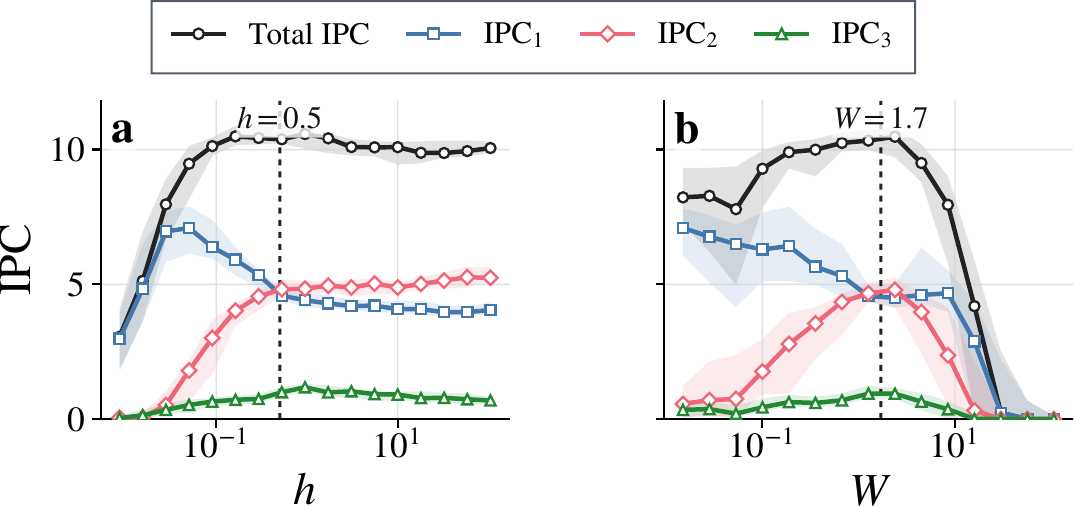}
    \caption{Total IPC and $\mathrm{IPC}_n$ for $n=1,2,3$ against Hamiltonian parameters. \textbf{a}: IPC results across a sweep of $h$ values for $W=0$. \textbf{b}: IPC results across a sweep of $W$ values for $h=0.03$.}
    \label{fig:phase3}
\end{figure}
\begin{figure*}
    \centering
        \includegraphics[width=0.9\linewidth]{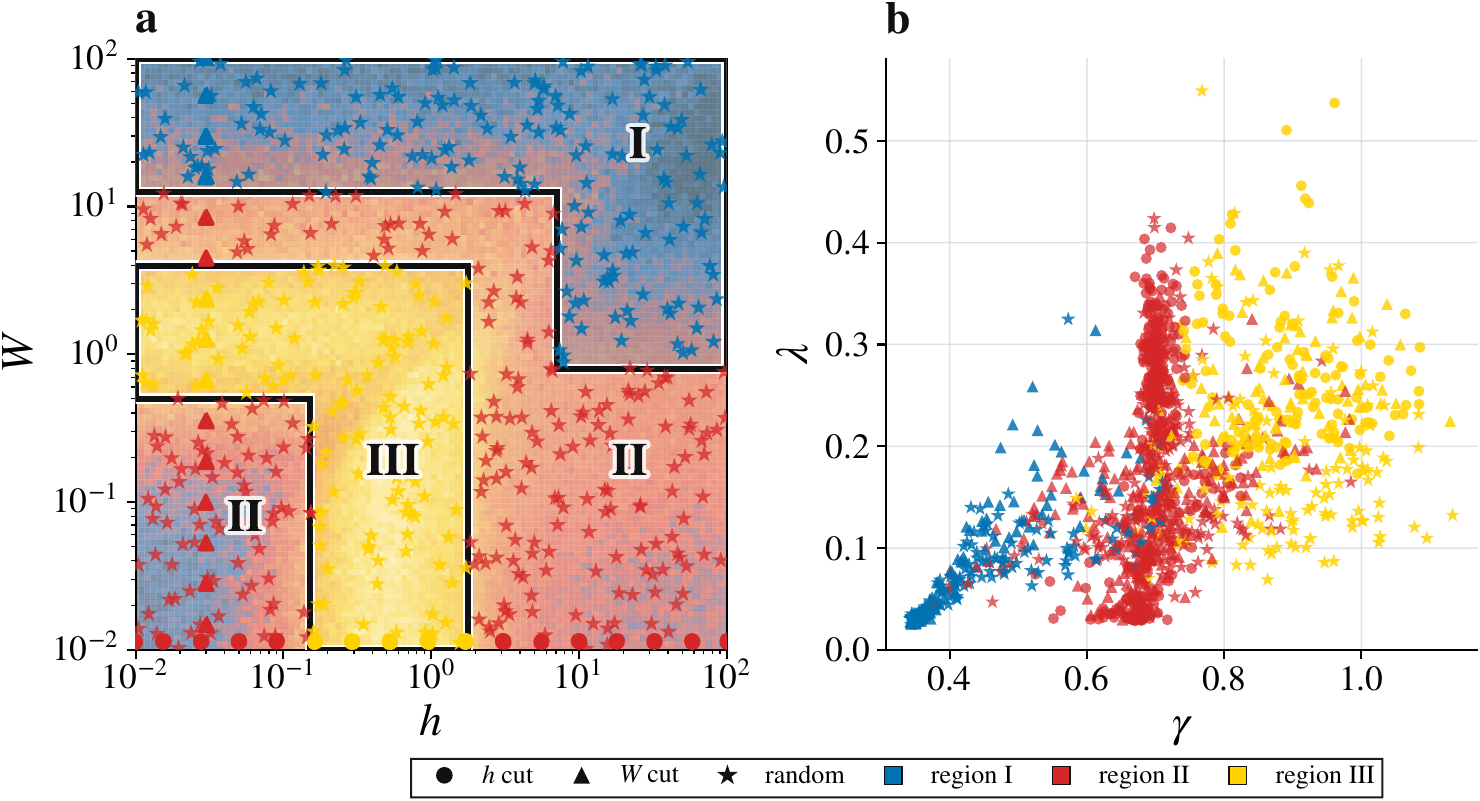}
    \caption{\textbf{a}: The phase diagram from \cref{fig:phase}, heuristically split into three regions, with markers indicating $(h,W)$ values used in the IPC results found in \cref{fig:ipc}. \textbf{b}: The $(\gamma, \lambda)$ values used for the IPC results, coloured to indicate which region of the phase diagram in \textbf{a} they come from.}
    \label{fig:phase2}
\end{figure*}
In order to calculate the QRC performances shown in \cref{sec:perform}, we run the protocol with uniform random inputs from $[0,1]$ with $1000$ washout steps, $2000$ training steps, and then test the performance over a subsequent $2000$ steps, and a least-squares cost function for the training described in \cref{sec:train} is used.

Performance for the individual tasks that make up the IPC are calculated by using squared Pearson correlation coefficients (PCCs). If the vector $\mathbf y$ consists of the values of a particular target function over the test cycles and $\mathbf x$  consists of the corresponding outputs of the trained linear readout layer described in \cref{sec:train}, the square PCC is given by:
\begin{equation}
   \mathrm{Pearson}(\mathbf x, \mathbf y)^2 = \frac{\mathrm{Cov}[\mathbf x, \mathbf y]^2}{\mathrm{Var}[\mathbf x]\mathrm{Var}[\mathbf y]}. 
\end{equation}
In order to reduce the effects of statistical noise, we employ a cut-off of $\epsilon=0.02$, so that if the square PCC of a specific target function is less than $\epsilon$, we set it to zero. We find that none of the target functions that make up $\mathrm{IPC}_n$ for $n>4$ pass this filter, and so the results for the total $\mathrm{IPC}$ in \cref{fig:ipc} have contributions from $\mathrm{IPC}_n$ for $1\leq n\leq 4$. We plot the IPC across $h$ and $W$ sweeps in \cref{fig:phase3}

The regions of high performance shown encircled in \cref{fig:ipc} are chosen in the following way. First, the $(\gamma, \lambda)$ space is coarse grained, and the square elements that have a higher median value than $65\%$ of the others are chosen. Isolated points are then excluded, and a smoothened boundary is drawn around the remaining region.

The path showing performance for reconstruction of the individual linear targets $y_\tau = \ell_1(s_{t-\tau}) = \sqrt{3}(2s_{t-\tau}-1)$ with increasing $\tau$ is calculated by taking the top $30$ performing points in the $\gamma$-$\lambda$ phase diagram for a particular target function $y_\tau$, and calculating its median position.

The IPC results presented in \cref{fig:ipc} show a clustering of points into a central vertical column, with two less defined clusters on either side. In order to analyse this behaviour, in \cref{fig:phase2} \textbf{a} we heuristically split the phase diagram into three regions, and colour the markers appropriately in panel \textbf{b}. We note that the localised regime, indicated by region I, yields the clustering of points in the lower left (weak scrambling, slow decay) of panel \textbf{b}. Region III, the region of `maximum chaos' as indicated by the $\langle r\rangle$ values in \cref{fig:phase}, instead explains the majority of the points on the right of the vertical column in \cref{fig:phase2}. Finally, the remainder of the phase diagram, indicated as region II, mainly describes the aforementioned vertical column of points. In this region, it seems that scrambling strength is mostly squeezed as a result of this column of points, and moving around region II serves mainly to tune the memory decay strength $\lambda$.

One final thing to note is that these regions are related to, but not solely determined by $\langle r\rangle$ values. For example, the lower left rectangle in \cref{fig:phase2} shares similar $\langle r \rangle$ values with region I, but is excluded as it contributes to the lower part of the vertical column (red) seen in panel \textbf{b}.

\end{document}